\documentclass[manuscript]{aastex}
\usepackage{natbib}
\usepackage{fancyvrb}
\usepackage{longtable}
\usepackage{epsfig}
\usepackage{fmtcount}
\usepackage{listings}
\usepackage{rotating}
\usepackage{amssymb}
\usepackage{enumerate}

\newcommand{\degree}{\ifmmode {^{\circ}} \else {$^{\circ}$} \fi}
\newcommand{\degrees}{\ifmmode {^{\circ}} \else {$^{\circ}$} \fi}

\newcommand{\unit}[1]{\ifmmode {\rm\ #1\,} \else {$\rm #1$} \fi}
\newcommand{\quarter}{\ifmmode {\frac{1}{4}} \else {$\frac{1}{4}$} \fi}

\newcommand{\kg}{\unit{kg}}
\newcommand{\J}{\unit{J}}

\newcommand{\us}{\unit{\mu s}}

\newcommand{\K}{\unit{K}}

\newcommand{\jyus}{\unit{Jy\ \mu s}}

\newcommand{\tten}[1]{\ifmmode {\times 10^{#1}} \else {$\times 10^{#1}$} \fi}
\newcommand{\tentothe}[1]{\ifmmode {10^{#1}} \else {$10^{#1}$} \fi}

\newcommand{\pcyr}{\unit{pc^{-3} yr^{-1}}}
\newcommand{\pccm}{\unit{pc\ cm^{-3}}}

\newcommand{\doublet}{\ifmmode {\lambda\lambda} \else {$\lambda\lambda$} \fi}
\newcommand{\singlet}{\ifmmode {\lambda} \else {$\lambda$} \fi}

\newcommand{\DM}{\mathrm{DM}}

\shorttitle{Astropulse: A Transient Radio Survey}
\shortauthors{Von Korff et al.}

\begin{document}

\VerbatimFootnotes

\title{Astropulse: A Search for Microsecond Transient Radio Signals Using Distributed Computing.  I. Methodology}

\author{J. Von Korff \altaffilmark{1} \altaffilmark{2} \altaffilmark{3},
        P. Demorest \altaffilmark{4},
        E. Heien \altaffilmark{1} \altaffilmark{5},
        E. Korpela \altaffilmark{1},
        D. Werthimer \altaffilmark{1},
        J. Cobb \altaffilmark{1},
        M. Lebofsky \altaffilmark{1},
        D. Anderson \altaffilmark{1},
        B. Bankay \altaffilmark{1},
        A. Siemion \altaffilmark{1}}
\altaffiltext{1}{Space Sciences Lab, University of California, Berkeley, CA 94720}
\altaffiltext{2}{Physics department, Georgia State University, Atlanta, GA 30303}
\altaffiltext{3}{vonkorff@gmail.com}
\altaffiltext{4}{National Radio Astronomy Observatory, Charlottesville, VA 22903}
\altaffiltext{5}{Graduate School of Information Science and Technology, Osaka University}

\begin{abstract}
We are performing a transient, microsecond timescale radio sky survey, called ``Astropulse,'' using the Arecibo telescope.  Astropulse searches for brief ($0.4 \us$ to $204.8 \us$), wideband (relative to its $2.5$ MHz bandwidth) radio pulses centered at $1{,}420$ MHz.
Astropulse is a commensal (piggyback) survey, and scans the sky between declinations of $-1.33$ and $38.03$ degrees.  We obtained $1{,}540$ hours of data in each of $7$ beams of the ALFA receiver, with $2$ polarizations per beam.  The data are $1$-bit complex sampled at the Nyquist limit of $0.4 \us$ per sample.
Examination of timescales on the order of microseconds is possible because we used coherent dedispersion, a technique that has frequently been used for targeted observations, but has never been associated with a radio sky survey.  The more usual technique, incoherent dedispersion, cannot resolve signals below a minimum timescale which depends on the signal's dispersion measure and frequency.  However, coherent dedispersion requires more intensive computation than incoherent dedispersion.
The required processing power was provided by BOINC, the Berkeley Open Infrastructure for Network Computing.
BOINC is a distributed computing system, allowing us to utilize hundreds of thousands of volunteers' computers to perform the necessary calculations for coherent dedispersion.
Astrophysical events that might produce brief radio pulses include giant pulses from pulsars, RRATs, exploding primordial black holes, or new sources yet to be imagined.  Radio frequency interference (RFI) and noise contaminate the data; these are mitigated by a number of techniques including multi-polarization correlation, DM repetition detection, and frequency profiling.
\end{abstract}

\keywords{radio continuum: general --- extraterrestrial intelligence --- pulsars: general --- black hole physics --- cosmology: early universe}

{\it Facility:} \facility{Arecibo(ALFA)}

\section{Introduction}

% This section should concern theory of black holes, pulsars, etc.

%%%%%%%%%%%%%%%%%%%%%%%%%%%%%%%%%%%%%%%%%%%%%%%%%%%%%%%%%%%%%%%%%%%%%%%%%%%%%%%%%%%%%%%%%%%%%%%%%%%%%%%%%%%%%%%%%%

\subsection{Scientific motivation}

This is an exciting time in the field of transient astronomy, both in the radio and in other parts of the spectrum.  
Improving technology allows astronomers to perform fast followups of transient events, store extensive digital records of
observations, and run processor-intensive algorithms on data in real time.  These advances make possible instruments that examine optical afterglows of gamma-ray bursts \citep{ves05} or neutrino sources \citep{kow07}.  High resolution digital images can be recorded and stored quickly using current \citep{kai04} and planned technology \citep{ive08}.  In the radio, astronomers search for transients such as orphan GRB afterglows \citep{lev02} or radio bursts of unknown origin \citep{kat03}.

Our project, called ``Astropulse,'' searches for brief, wideband radio pulses on timescales of microseconds to milliseconds, and surveys the entire sky visible from Arecibo Observatory.  The idea of a short-timescale radio observation is not new.  Other experiments are well-suited for detecting radio pulses on a microsecond timescale, or even much shorter scales.  However, these observations are directed; they examine known phenomena.  For instance, such an experiment might record the nanosecond structure of the signals from the Crab pulsar.  And of course the idea of a radio survey is not new.  Other experiments perform surveys for radio pulses over large regions of the sky.  However, these
observations examine $50 \us$ timescales or longer.  Astropulse is the first radio survey for transient phenomena with microsecond resolution.

This project is made possible by Astropulse's access to unprecedented processing power, using the distributed computing technique.  Because the interstellar medium disperses radio signals, all of our data must be dedispersed.  We send our data to volunteers, who perform coherent dedispersion \citep{lor05, han75} using their own computers.  Then they send the results of this computation back to us, informing us whether they detected a signal, and reporting that signal's dispersion measure, power, and other parameters.
Astropulse is processor intensive because we must perform coherent dedispersion, whereas other surveys perform incoherent dedispersion.  Coherent dedispersion is necessary to resolve structures below $50\;\mu$s or so, depending on the dispersion measure.

We are not committed to detecting any particular astrophysical source; rather,
we are motivated by our ability to examine an unexplored region of parameter
space.  However, we consider that we might detect evaporating primordial black
holes, millisecond (or faster) pulsars, or RRATs \citep[rotating radio transients,][]{mcl06}.  We will consider each of these possibilities in turn.  We could potentially detect pulsed communications from extraterrestrial civilizations, though we do not discuss this possibility herein.

%%%%%%%%%%%%%%%%%%%%%%%%%%%%%%%%%%%%%%%%%%%%%%%%%%%%%%%%%%%%%%%%%%%%%%%%%%%%%%%%%%%%%%%%%%%%%%%%%%%%%%%%%%%%%%%%%%

\subsection{Black holes}

\subsubsection{Hawking radiation}

It was proposed by \citet{haw74} that a black hole of mass $M$ emits radiation like a black body whose temperature is given by the following relation:

\begin{equation} \label{eq:TandM}
T_{BH} = \frac{\hbar c^3}{8\pi k G M}
  = 10^{-6} \left( \frac{M_{\odot}}{M} \right) \K.
\end{equation}

The radiant energy comes directly from the black hole's mass, and as a result,
it is losing mass at a rate $\dot{M} \propto -M^{-2}$.  
Because the black hole radiates more power as it shrinks, we expect a burst of energy in the last moments of the black hole's life.  
One can make different assumptions about the energy distribution of the radiation from a black hole evaporation \citep{car76}.  For a ``hard'' equation of state, with an adiabatic index $\Gamma > \frac{6}{5}$, the radiation does not reach thermal equlibrium.  The standard model falls into this category, and it would assume that the radiation behaves as a relativistic ideal gas, $\Gamma = \frac{4}{3}$.  In this case, the final explosion of the black hole lasts on the order of seconds.  However, for a ``soft'' equation of state, as proposed by \citet{hag65}, $\Gamma$ could be much smaller.  In this case, the explosion might happen in ${10}^{-7}$ seconds or less.  Astropulse is ideally suited for detecting such fast explosions.

We can integrate the radiant energy to find the total lifetime of the black hole, demonstrating that if the black hole is exploding now, it must have been created at a mass of ${10}^{12} \kg$ or less.
Such a small black hole cannot have been born from a star, and would have to be created in the big bang \citep{haw71}, from
``density perturbations in the early universe'' \citep{mac90}.  

\subsubsection{Electromagnetic pulses}

The total amount of energy released in the last second of the black
hole's life is about $10^{23} \J$.  Most previous studies have attempted
to detect this energy in the cosmic gamma ray background \citep{rai05}.
But \citet{ree77}
suggested that some of this energy could be converted into a radio
pulse.  The idea is that as the black hole shrinks and becomes hotter, it
starts radiating more and more massive particles, including
electrons and positrons (due to pair production at the event horizon),
but later, heavier particles as well.  This forms a plasma fireball
expanding around the black hole.  As this conducting shell expands into
the ambient magnetic field, it pushes the field out of the way, creating
an electromagnetic pulse.  Rees argued that for a magnetic field $B$ around $5 \times 10^{-6}$ Gauss and a critical
mass of $\sim 2 \times 10^{11}$ g, a radio pulse detectable
in the $21$ cm band is plausible.

An observation of these pulses would be a very significant confirmation
of both Hawking radiation and the existence of primordial black holes (PBHs).
At the very least, we can put a limit on the possible
maximum density of evaporating black holes in the universe, if we make some
assumptions about their distribution, and contingent on the assumption that they produce radio pulses.  
This information would be relevant to
cosmological models describing the big bang.

Some groups have searched for these PBHs in the radio, but many researchers have looked for gamma-ray emission instead \citep{ukw10}.  Radio and gamma-ray surveys make very different assumptions about the PBHs' evaporation time, so their results are difficult or impossible to compare in a meaningful way.  Ukwatta et al. describe PBH explosions as having timescales of seconds or minutes, whereas Astropulse is looking for microsecond pulses.  

%%%%%%%%%%%%%%%%%%%%%%%%%%%%%%%%%%%%%%%%%%%%%%%%%%%%%%%%%%%%%%%%%%%%%%%%%%%%%%%%%%%%%%%%%%%%%%%%%%%%%%%%%%%%%%%%%%

%%% entropy
%%% temperature
%%% power output
%%% 
%% note about negative specific heat?
%% black holes: two or three sizes have been observed
%%% solar mass: how they are created
%%% supermassive: how they are created
%%% intermediate mass black holes: have they been observed?
%%

\subsection{Other sources}

%% Handbook of Pulsar Astronomy - Lorimer & Kramer
%% Pulsar Astronomy - Lyne & Graham-Smith
%% probably Lorimer's is better.

%% Papers:
%% probably want a book on pulsars.
%%
%% Neutron Stars and Pulsars - Becker, Werner (Ed.)  Astrophysics and Space Sci Library, vol 357
%% Pulsar Astronomy, 3rd Edition (Cambridge Astrophysics) by Andrew G. Lyne (Author), Francis Graham-Smith (Author)
%% Handbook of Pulsar Astronomy (Cambridge Observing Handbooks for Research Astronomers) (Hardcover)
%%   by D. R. Lorimer (Author), M. Kramer (Author)

Astropulse might also detect RRATs \citep{mcl06} or repeating or giant pulses from pulsars.  Of these possibilities, giant pulses are the most likely.
Astropulse is optimized for pulses of $200\;\mu s$ or less, but its sensitivity relative to
other surveys is best at short timescales, around $0.4$ to $1.6\;\mu s$.  These timescales are much too short for repeating pulses from a pulsar \citep{lat90}, even a millisecond pulsar with a small beam opening angle half-width, or from a RRAT. 

Giant pulses, on the other hand, can have very short timescales suitable for detection by Astropulse.
The Crab's giant pulses have structure ranging
 from a few nanoseconds in duration at $2\:\jyus$ \citep{han03} to $64\:\us$ or more at
$1{,}000$ to $10{,}000\: \jyus$, with a typical
pulse duration of a few microseconds \citep{pop07}.  (See Section~\ref{sec:overviewalgorithm} for a discussion of the $\jyus$ unit.)

% Prior work

%% Papers: find them on my USB?  Or do I have them on my computer?
%%
%% Discuss relation between limit on black hole evaporation vs. bandwidth, system temperature, gain,
%% observing time x # of beams x # of polarizations, beam width.
%% Is it better to have a single large dish, or many smaller dishes, for a given area?
%% Table: show all these numbers for prior experiments

% Why this is new

%% Papers: reference Kraus?
%%
%% Compare coherent and incoherent dedispersion
%% Formula for incoherent dedispersion - tradeoff between narrower and wider subbands

%% 

%% Geoff says:
%% Stellar Phenomena?
%% Other sub-second transients?
%% GRB coherent pulses?
%% NS-NS mergers

\section{Telescope and instrumentation} \label{sec:telescope}

\subsection{Sky coverage}

Arecibo Observatory scans approximately one third of the sky, between 
declinations of $-1.33$ and $38.03$ degrees.  Because of this, Astropulse cannot 
see the galactic center (around $-29^\circ$ dec) but can see $452$ out of the $1826$ 
pulsars in the ATNF pulsar database\footnote{\verb,http://www.atnf.csiro.au/research/pulsar/psrcat/,, as of 6/29/2009}, including the Crab.  Astropulse is a commensal survey; this means that other surveys control the telescope pointing, but allow Astropulse to collect data at all times.  Our partner surveys include GALFA (Galactic ALFA), discussed in \citet{peek08, sta06} and PALFA (Pulsar ALFA), discussed in \citet{cor08}.  Our group also operates SETI@home, another commensal radio survey, and the two projects use the same data: a 1-bit complex sampled $2.5$ MHz bandwidth centered at $1420$ MHz.  To date, we have observed for $1{,}540$ hours with each of the $7$ beams (and $2$ linear polarizations per beam), for a total of $21{,}600$ hours of observation time.  We have been taking our primary set of data using the ALFA receiver from September 2006 until May 2010, for a total of $3.7$ years.  This implies that we have had $1/21$ of all possible observation time during those years.  Since a good deal of Arecibo's time is dedicated to non-astronomical purposes, such as ionospheric science, our fraction of astronomy time is significantly larger than $1/21$.

\subsection{ALFA receiver}

The ALFA (Arecibo L-band Feed Array) receiver has $7$ dual-polarization beams on the sky arranged in a hexagonal pattern, each with a $3.5'$ beamwidth.
The central beam has a gain of $11$ K / Jy, and the other beams have $8.5$ K / Jy  
\footnote{\verb,http://www.naic.edu/alfa/gen_info/info_obs.shtml,}.  The system temperature is $30$ K.  The $6$ peripheral beam pointings differ from the central beam by a maximum of $6.4'$. 

\subsection{Downconverter}

Multiple experiments use the signal from the ALFA receiver, so we split the signal 
using an IF splitter.  These $14$ signals are attenuated by $6$ to 
$13$ decibels for purposes of level-matching, and then enter our multibeam quadrature 
baseband downconverter.  Downconversion involves complex multiplication, resulting in $14$ complex or $28$ real channels.
%\begin{figure}[!h] 
%\includegraphics[width = 6.0in]{downconverter.eps}
%\caption[Astropulse backend]{Astropulse / SETI@home backend, including attenuators, downconverter,
%\label{fig:downconverter}
%and Dell PC.  The DDA cards are labelled ``EDTCD60.''}
%\end{figure}

\subsection{Data recorder}

These $28$ real channels are digitized with $1$ bit precision using comparators, and the resulting digital signals (the signs of the $28$ voltages) are directed through ribbon cables to Digital Data Acquisition (DDA) cards on a PC,  
which is running our software that acquires and writes the data to disk.  
In addition, 
this software collects telescope coordinates from the Arecibo telescope's data broadcast network,
SCRAMnet.  The coordinates consist of the Right Ascension (RA)  and Declination (Dec) to which the
 telescope is currently pointing, as well as the time for which that RA and Dec are valid.

The data files are stored on a hot swappable SATA drive, which fills up in $14$ to $20$ hours of 
observation time.  Since we are taking data $1/21$ of the time, we must swap out the SATA drive about once
per two weeks.
When enough drives have been collected,
 the Arecibo staff ship the drives to us at Space Sciences Lab, UC Berkeley.
We use $20$ SATA drives in all, each of which holds $500$ or $750$ GB.  We also send a backup copy of each file to NERSC, the National Energy Research Scientific Computing Center.  This ensures that we can retrieve the data at any time.
In all, we have taken over $48$ TB of data from ALFA multibeam.
We need a large
(6 TB) disk array at Berkeley to buffer the data before it is sent to volunteers.
The volunteers' PCs then process the data and send the results back to Berkeley.  
For a discussion of data processing after this point, see Section~\ref{boinc} on BOINC, and
Section~\ref{sec:detection} on the dedispersion algorithm for the volunteers' client program.

\section{Pulse detection : thresholds and dedispersion} \label{sec:detection}

\subsection{Overview} \label{sec:overviewalgorithm}

The primary function of the Astropulse program is to dedisperse potential pulsed signals,
then determine whether the dedispersed pulse surpasses an appropriate power threshold.  Pulses with sufficient power are recorded as ``candidate pulses'' in the Astropulse database.  We will discuss
the theory behind dedispersion and then the methods we use to select the
thresholds.  Then we will calculate the nominal sensitivity of Astropulse in $\jyus$, a unit of ``pulse area.''

The $\jyus$ unit refers to the pulse's flux density (in Jy) integrated over its
duration (in $\us$.)  It is called an ``area'' because it is calculated using
this integral, which is the area under a curve.  Although the unit of flux
density (Jy) is a more conventional measure of sensitivity, the $\jyus$ is more
meaningful in our case, because we are attempting to detect unresolved pulses.  For example, consider two pulses; one is $500$ Jy and lasts $0.2\; \us$, and the other is $1{,}000$ Jy and lasts $0.1\; \us$.  When these pulses are dispersed, they will be similar in appearance; Astropulse cannot distinguish between them because their dedispersed durations are shorter than Astropulse's time resolution.  But Astropulse can determine that both pulses are $100\; \jyus$.  

Note that when we describe a measured pulse's apparent area in $\jyus$, the
actual pulse area may be different depending on any contributions to the system
temperature.  We assume a particular minimal system temperature ($30$ K) for the ALFA receiver, whereas we might have a different effective system temperature when looking at the Crab nebula.

% Coherent de-dispersion of pulses that traveled through the ISM

\subsection{Dedispersion} \label{dedispersion}

Between a radio pulse's source (i.e. black hole, pulsar, or ET) and our detector, the pulse is dispersed as it travels through the Interstellar Medium (ISM).  According to \citet{lor05}, the relative time delay for frequency $\nu$ is given by:

\begin{eqnarray}
t(\nu) = \mathcal{D} \times \mathrm{DM}/\nu^2 \label{eq:tofnu} \\
\mathrm{DM} = \int n_e d\ell
\end{eqnarray}

where the integral is over the distance to the source of the pulse, $n_e$ is the electron density, and $\mathcal{D}$ is equal to $4.15 \times {10}^3 \; \mathrm{MHz}^2\; \mathrm{pc}^{-1}\; \mathrm{cm}^3\; \mathrm{s}$.

A useful estimate for the dispersion measure weighted mean electron density in our Galaxy is
 $n_e = 0.03$ cm$^{-3}$ \citep{gue73}.  
% Consider, for instance, the Crab pulsar, at DM = $56.8 \pccm$.  The difference, $\Delta \tau$, between the high and low frequency time delay observed by Astropulse is $0.411$ s.  

Astropulse loops through the data at several nested levels, and considers DMs ranging from $-830 \pccm$ to $-49.5 \pccm$ and from $49.5 \pccm$ to $830 \pccm$.  We chose the lower limit, $49.5 \pccm$, for two reasons.  First, we found that our sensitivity diminishes at low dispersion measures due to the effects of one-bit digitization.  For instance, a (hypothetical) very strong, undispersed, $2 \us$ signal would be undetectable, since its signature in our data would only be five samples long, and each sample carries only one bit of information.  Therefore, we can only detect dispersed pulses.  Second, local interference at Arecibo Observatory is stronger at low dispersion measures.  We tested the first effect by inserting simulated pulses into our detection algorithm, and the second effect by examining data from the telescope.  In this way, we empirically determined a lower limit for our dispersion measure.  The upper limit of $830 \pccm$ was selected so that approximately half of the volume of the Galactic plane would be visible to our search according to the Galactic maps in \citet{cor03}.

Astropulse considers pulses of widths ranging from $0.4 \us$ (a single sample) to $204.8 \us$.  The larger widths are tested by summing $2^\ell$ adjacent samples after dedispersion, where $0 \le \ell \le 9$, and $\ell$ takes integer values.

\subsubsection{Incoherent dedispersion and its limitations} \label{sec:incohlimit}

We have two choices for our methodology: coherent dedispersion and incoherent 
dedispersion.  Astropulse uses coherent dedispersion, whereas other radio surveys
 use incoherent dedispersion.   Incoherent dedispersion is much more computationally efficient, and for 
longer timescales it's almost as good as coherent dedispersion.  However, as we 
will see, Astropulse would be unable to examine the $0.4\; \mu s$ timescale 
without coherent dedispersion.

Incoherent dedispersion means that the signal's power spectrum is calculated, and the power vs. time of each sub-band is analyzed. 
The method is called ``incoherent''
 for this reason -- the phase information about individual frequencies is lost;
only the total power of each sub-band at each time is retained.  Next, the sub-bands are realigned
at all possible dispersion measures, in an effort to find one DM at which the
components align to produce a large power in a short period of time. 

However, incoherent dedispersion is limited in 
two ways.  First, the goal of recording power vs. time makes sense only on a timescale 
greater than 

\begin{equation}
dt_1 = \frac{1}{d \nu}, \label{eq:dt1}
\end{equation}

where $d \nu$ is the width of each sub-band.  
This is because of time-frequency uncertainty.

Second, in each sub-band the pulse is dispersed by some amount $dt_2(\nu)$, which we can find using Equation~\ref{eq:tofnu}.

\begin{eqnarray}
dt_2(\nu) & = & t(\nu) - t(\nu + d\nu) = 
\frac{\mathcal{D} \cdot \mathrm{DM}}{\nu^2} 
- \frac{\mathcal{D} \cdot \mathrm{DM}}{(\nu + d\nu)^2} \\
& = & \frac{\mathcal{D} \cdot \mathrm{DM}}{\nu^2} (1 - \frac{\nu^2}{(\nu + d \nu)^2}) \\
& = & \frac{\mathcal{D} \cdot \mathrm{DM}}{\nu^2} (1 - (\frac{\nu}{\nu + d \nu})^2) \\
& = & \frac{\mathcal{D} \cdot \mathrm{DM}}{\nu^2} (1 - (1 - \frac{d \nu}{\nu + d \nu})^2) \label{eq:dt2band}
\end{eqnarray}

Under the assumption that incoherent dedispersion divides the band up into many small pieces, $d\nu \ll \Delta \nu$, or that the bandwidth is much smaller than the frequency, $\Delta \nu \ll \nu$, we obtain $d \nu \ll \nu$.  Then we can approximate $\frac{d\nu}{\nu} \ll 1$ in Equation~\ref{eq:dt2band}, so that the equation becomes:

\begin{eqnarray}
dt_2(\nu) & \approx & \frac{\mathcal{D} \cdot \mathrm{DM}}{\nu^2} (1 - (1 - 2 \frac{d \nu}{\nu})) \\
& = & 2 (\mathcal{D} \cdot \mathrm{DM}) \frac{d \nu}{\nu^3} \label{eq:dt2}
\end{eqnarray}

Incoherent dedispersion cannot probe timescales shorter than $dt_1$ or $dt_2(\nu)$.  Since $dt_1$ is inversely related to $d\nu$ while $dt_2(\nu)$ is proportional to $d\nu$, we can optimize the time resolution by selecting the optimal value of $d\nu$.  This will be the value of $d\nu$ such that $dt_1 = dt_2(\nu)$, that is:

\begin{eqnarray} \label{eq:dtsqrt}
2 (\mathcal{D} \cdot \mathrm{DM})\; \frac{d \nu}{\nu^3} & = & \frac{1}{d \nu} \\
{d \nu}^2 & = & \frac{\nu^3}{2 \mathcal{D} \cdot \mathrm{DM}} \\
d \nu & = & \sqrt{\frac{\nu^3}{2 \mathcal{D} \cdot \mathrm{DM}}} \\
dt(\nu) & = & \sqrt{\frac{2 \mathcal{D} \cdot \mathrm{DM}}{\nu^3}} \label{eq:dtcompromise}
\end{eqnarray}

where the last step used Equation~\ref{eq:dt1}.  For the Crab pulsar, the dispersion measure is $56.8 \pccm$ \citep{sal99}, and we are observing at a frequency of $1420$ MHz.  Then we can substitute $\mathcal{D} = 4.15 \times {10}^3\;\mathrm{MHz}^2\;(\pccm)^{-1}\;\mathrm{s}$, $\mathrm{DM} = 56.8 \pccm \times \frac{DM}{56.8 \pccm}$, and $\frac{1}{\nu} = \frac{1420\;\mathrm{MHz}}{\nu} \times \frac{1}{1420\;\mathrm{MHz}}$ to obtain:

\begin{eqnarray}
dt(\nu) & = & \sqrt{(8.3 \times 10^3\;\mathrm{MHz}^2\;(\pccm)^{-1}\;\mathrm{s})\;\times 56.8 \pccm} \nonumber \\
& \times & \sqrt{\frac{\mathrm{ DM }}{56.8 \pccm}\;(\frac{1420\;\mathrm{MHz}}{\nu})^3 \times (\frac{1}{1420\;\mathrm{MHz}})^3} \label{eq:DtauLyneGSc} \\
& = & 12.8\;\us\;(\frac{DM}{56.8 \pccm})^{0.5} (\frac{\nu}{1420\;\mathrm{MHz}})^{-1.5}. \label{eq:minresolution}
\end{eqnarray}

So for the Crab pulsar, this is a limit of $12.8\; \us$, or $32$ samples at $1420$ MHz.  For a 
more distant source, the limit might be as much as $50\; \mu s$, or $124$ samples.  (Astropulse considers sources with a DM as high as $830 \pccm$.)

\subsubsection{Coherent dedispersion as deconvolution}

Coherent dedispersion \citep{lor05, han75} is an alternative technique that allows better time
resolution by performing the mathematical inverse of the ISM's dispersion operation.  Coherent dedispersion deals with amplitude rather than power,
preserving phase information.  In the absence of noise or scattering, and given precise knowledge of the pulse's dispersion measure, coherent dedispersion would reconstruct the original pulse exactly.  We need to analyze the mathematical operation corresponding to dispersion in order to find its inverse.

If $F(n)$ is the original pulse as a function of sample number $n$, suppose $D[F]$ is the dispersed pulse.  Then, relying on the time translation invariance and linearity of the dispersion operator, we can show that $D$ is just a convolution.  In particular, $D[F] = F * D[\delta]$, where $*$ is convolution and $\delta$ is the discrete $\delta$ function.  Finally, the convolution theorem for Fourier transforms gives us:

\begin{equation} \label{eq:divbydft}
F  =  \mathrm{DFT}^{-1}(\frac{\mathrm{DFT}(D[F])}{\mathrm{DFT}(D[\delta])}).
\end{equation}

Equation~\ref{eq:divbydft} gives us a fast method for dedispersing a pulse $D[F]$,
 obtaining the original pulse $F$.  This method is fast because the Fast 
Fourier Transform (FFT) algorithm is fast, taking time $O(N \log N)$ to Fourier transform $N$ samples of data.  We can use this fact to estimate the run time of Astropulse's dedispersion algorithm.  Astropulse has to dedisperse each set of $N$ samples many times, since one dedispersion must be performed for every dispersion measure.  If $M$ is the number of dispersion measures to be tested, then $N$ samples can be dedispersed in time $O(M N \log N)$.  Furthermore, Astropulse must operate on a long stream of data, of length $L$, which is much longer than the length $N$ of a single Fourier transfom.  Therefore, the total time required is $O(M L \log N)$.

A similar calculation suggests that incoherent dedispersion would be faster.  This is partly because incoherent dedispersion performs fewer tests.  Incoherent dedispersion cannot test as many dispersion measures as coherent dedispersion does, because its time resolution is imperfect and it cannot always distinguish between different dispersion measures.  To test a particular dispersion measure, an algorithm must target a particular time delay between the minimum and maximum frequency in the band.  This time delay cannot be determined more accurately than the time resolution $dt$ (Equation~\ref{eq:minresolution}) resulting from incoherent dedispersion.  If $dt$ corresponds to $n$ samples, then incoherent dedispersion can test only $\frac{1}{n}$ as many dispersion measures as coherent dedispersion.  In other words, with incoherent dedispersion, we would test not $M$, but $M / n$ dispersion measures.

With incoherent dedispersion, we must process $L$ samples for each dispersion measure, so we require time $O(M L / n)$.  (Some additional time is required to Fourier transform the data into a power spectrum, but this process is not dominant.)  So the time ratio between coherent and incoherent dedispersion is $O(M L \log N / (M L / n)) = O(n \log N)$.  In our case, the time resolution $dt$ will vary by dispersion measure, but assuming a value of $20 \us$, corresponding to a DM of $139 \pccm$, and a sample duration of $0.4 \us$, we obtain $n = 50$.  Then $\log N = 15$, so $n \log N = 750$, a large number.  Therefore, we expect that incoherent dedispersion is substantially faster than coherent dedispersion, as long as both methods are applicable.  Coherent dedispersion is useful in situations where a very short time resolution is required, shorter than that permitted by incoherent dedispersion.

\subsubsection{Nonlinear chirp function} \label{sec:nonlinear}

The ``chirp'' function, $\mathrm{DFT}(D[\delta])$, plays an important role in Equation~\ref{eq:divbydft}.  
The function $D[\delta]$ represents a dispersed delta function, 
and the $\mathrm{DFT}$ transforms it into the frequency domain.  
In other words, we imagine a brief, strong pulse emitted by an 
astrophysical source.  The pulse is dispersed by its passage through the 
interstellar medium, resulting in a chirp function that is spread out in time.
(The name ``chirp'' is meant to suggest a sound with changing frequency, just 
as the chirp signal has a changing radio frequency.)  We can then compute the functional form of this dispersed signal in the frequency domain.

This computation has been performed by \cite{han75}.  They write the chirp function in the frequency domain as:

\begin{equation} H_+(f + f_0) = \exp \left[ \frac{i 2 \pi D f^2}{f_0^2 (f_0 + f)} \right], f \ll f_0 \label{eq:nonlinearchirp} \end{equation}

where $H_+$ is the transfer function (which is our chirp function), $f$ is the frequency relative to band center, $f_0$ is the band center frequency, and $D$ is the dispersion coefficient corresponding to our $\mathcal{D} \times \mathrm{DM}$.  Hankins and Rickett also describe a linear approximation (not shown here), which they say is useful ``for many applications.''  However, a linear approximation is not valid for our application because the curvature of the pulse arrival time-frequency profile can be large compared to our $0.4 \us$ resolution.  The Hankins and Rickett approximation would be exactly correct for a time delay $t_\mathrm{lin}(\nu)$ that is linear in $\nu$.  The approximation fails when the difference between the time delay and the linear approximation to the time delay, $t(\nu) - t_\mathrm{lin}(\nu)$, is greater than the larger of the pulse duration or the sample duration.  To estimate this differential time delay, we take the second-order term in the Taylor expansion of $t(\nu)$ across the band: $\frac{d^2}{d\nu^2} t(\nu) (\Delta \nu)^2 = \frac{d^2}{d\nu^2} (\mathcal{D} \times \mathrm{DM} \times \nu^{-2}) (\Delta \nu)^2 = (\mathcal{D} \times \mathrm{DM}) 6 \nu_0^{-4} (\Delta \nu)^2 = 2 \us$, substituting the values for the Crab pulsar from Section~\ref{sec:incohlimit}.  This is bigger than our $0.4 \us$ time resolution, and it would be even larger at larger DMs, so we have chosen to use the exact (nonlinear) chirp function given by Equation~\ref{eq:nonlinearchirp}.

% Logic of the algorithm

% \subsection{Algorithm logic} \label{logic}

% \input{logic}

% Thresholds for noise rejection

\subsection{Thresholds for noise rejection} \label{thresholds}

In the absence of an astrophysical signal or radio frequency interference (RFI), we would detect only noise.  The noise would result in a random binary sequence in our data stream.  This noisy data can appear to contain pulses, simply by chance, with smaller noise pulses being more common and larger pulses less common.  Since very large pulses are quite rare, Astropulse can reject noise by searching for pulses whose power exceeds certain thresholds.  These thresholds can be calculated either experimentally or theoretically.  We'll start by finding the theoretical values, then point out some of the uncontrollable factors that make these values inaccurate, and finally we'll describe a Monte Carlo simulation method for calculating thresholds.  In this section, we will discuss only the rejection of noise and not RFI; the latter will be reserved for Section~\ref{sec:rfimitigation}.  In addition, the method discussed here will not suffice to reject all noise; a further method for noise rejection will be discussed in Section~\ref{sec:multipols}.

\subsubsection{Noise rejection thresholds: theory} \label{sec:singlepulsetheory}

We want to calculate the distribution (pdf) of the integrated noise power in $2^\ell$ samples, after
dedispersion.  (Here, the ``power'' refers to the absolute value of the square of the amplitude, where the undispersed time series has amplitudes $\pm 1$.)  We will perform some calculations and conclude that the dedispersed signal $f_d(t)$ is distributed like a complex Gaussian at each time $t$.

  First, we assume that the pre-dedispersion time series is pure white noise; that is, each bit of a two bit complex sample is
independently distributed with equal probability of a $0$ or $1$, so each $f(t)$ has equal
probability for $\pm 1 \pm i$.  (Note that this $f$ is discrete-valued with a discrete time argument, as opposed to the continuous-valued $f$ with continuous time argument described in Section~\ref{sec:nonlinear}.)  Then we
deconvolve this data by FFT.  In other words, 

\begin{equation}
\tilde{f}(k) = \frac{1}{\sqrt{N}} \sum_{t = 0}^{N-1} f(t) e^{- 2 \pi i k t / N}.
\end{equation}

The distribution of a single $\tilde{f}(k)$ is Gaussian (by the central limit theorem),
and to deduce its variance, we will find the variance of its real and complex components $\Re(\tilde{f}(k))$ and $\Im(\tilde{f}(k))$ independently:

\begin{eqnarray}
\Re(\tilde{f}(k)) & = & \frac{1}{\sqrt{N}} \sum_{t=0}^{N-1} \Re(f(t)) \cos(2 \pi k t / N) - \Im(f(t)) \sin(2 \pi k t / N) \\
\mathrm{Var}(\Re(\tilde{f}(k))) & = & (\frac{1}{\sqrt{N}})^2 \sum_{t=0}^{N-1} \Bigg( \mathrm{Var}(\Re(f(t))) \cos^2(2 \pi k t / N) \nonumber \\
      & & + \mathrm{Var}(\Im(f(t))) \sin^2(2 \pi k t / N) \Bigg) \\ 
& = & \frac{1}{N} \sum_{t=0}^{N-1} \cos^2(2 \pi k t / N) + \sin^2(2 \pi k t / N) \label{eq:variance}\\
& = & \frac{1}{N} \cdot N = 1
\end{eqnarray}

Equation~\ref{eq:variance} follows from the previous equation because $f(t)$ can only take the values $\pm 1 \pm i$, therefore its real and imaginary components each have a variance of $1$.

Therefore, the variance of the real component of $\tilde{f}(k)$ is $1$, and the same argument holds for the imaginary component.  Once we have obtained $\tilde{f}(k)$, the remaining steps in the dedispersion are to multiply by a frequency-domain chirp function, followed by Fourier transforming back to the time domain.

The frequency-domain chirp function has the form $e^{i \tilde{\theta}(k)}$ for some real phase $\tilde{\theta}$.  Since $\tilde{f}(k)$ is already a complex number with random phase, multiplication by another complex phase has no effect on the probability distribution of $\tilde{f}(k)$.  Finally, we run the inverse Fourier transform to obtain a dedispersed signal, $f_d(t)$.  Since $\mathrm{Var}(\Re(\tilde{f}(k))) = \mathrm{Var}(\Re(f(t))) = 1$, the same mathematical argument that we applied to the forward Fourier transform also applies to the inverse Fourier transform.  Then we find that the dedispersed signal $f_d(t)$ is again a complex Gaussian, such that $\Re(f_d(t))$ and $\Im(f_d(t))$ are Gaussians with variance 1.

The power in each sample after dedispersion will be distributed as $|\Re(f_d(t))|^2 + |\Im(f_d(t))|^2$, the sum
of the squares of two standard Gaussians.  This distribution is easily calculated; the 
joint probability distribution is:

\begin{eqnarray}
& & \frac{1}{\sqrt{2 \pi}} e^{\frac{-x^2}{2}} \cdot \frac{1}{\sqrt{2 \pi}} e^{\frac{-y^2}{2}} dx dy \\
& = & \frac{1}{2 \pi} e^{- \frac{r^2}{2}} r dr d\theta \\
& \rightarrow & e^{- \frac{r^2}{2}} r dr \\
& = & e^{-u} du.
\end{eqnarray}

where $u = \frac{x^2 + y^2}{2}$ is half the power in one sample, in the time domain.  Therefore,
half the power is exponentially distributed with mean $1$; or equivalently, the power is 
exponentially distributed with mean $2$.

In future calculations, we will normalize to half of this power, so that the average power per sample is $1$.

So for instance, if after dedispersion we find that a certain sample has a power $P$, we conclude that
only one in $e^P$ samples has a comparable power.  To ascertain how unlikely this is, we need to
calculate how many such samples we have examined over the entire course of the experiment.  This would be:

\begin{eqnarray}
% 7 \textrm{ beams } \times 2 \textrm{ polarizations } \times \frac{1}{3} \textrm{ on fraction } \times 3 \textrm{ years } \nonumber \\
% \times \; 1 \textrm{ workunit per } 13\; s \times 2^{25} \textrm{ samples per workunit } \times 14208 \textrm{ DMs } \nonumber \\
48 \textrm{ TB } \times 4 \cdot 10^{12} \textrm{ samples per TB } \times 14208 \textrm{ DMs } \nonumber \\
\times\; 2 \textrm{ DM signs } \nonumber \\
 = 5.45 \times 10^{18} = e^{43.1}.  \label{eq:countpulses}
\end{eqnarray}

So far, we have considered pulses that are one sample in width.  However, we are also searching for pulses of width $2, 4, 8, \ldots, 512$ samples.  To search for wider pulses, we sum the power over $2^\ell$ adjacent samples, for each integer $\ell$ between $0$ and $9$.  We will refer to this summation as ``co-adding,'' and each of the $10$ possible widths is a ``co-add.''  We compare this summed power to an appropriate threshold, which is larger at higher co-adds.  There are half as many potential pulses at each co-add, compared with the
previous co-add.  So Equation~\ref{eq:countpulses} underestimates the number of potential pulses by a factor of $1 + \frac{1}{2} + \frac{1}{4} + \ldots + \frac{1}{512} \approx 2$.)  

Therefore, with a threshold of $43.8$ for one-sample potential pulses, and appropriate thresholds for potential pulses of other sizes, we would rule out all but one noise event over the course of our entire observation history.  Thus, if Astropulse had no mechanism to remove noise other than to raise the detection threshold, our threshold would have to be quite high.  Fortunately, we do have an alternative mechanism; we discuss this issue further in Section~\ref{sec:multipols}.

Rather than raising the threshold to $43.8$, it seems more prudent to aim for one noise event to exceed threshold in each workunit -- a unit of data defined to be $13$ s for logistical reasons related to our data processing methods -- and sort out false pulses later.  (More than one pulse per workunit would be difficult to store in our database.)

In this case, we just want to find $C$, the number of samples we examine, multiplied by $2$ to account for co-adds.  This gives us the number of potential pulses (counting all co-adds) per workunit.

\begin{equation} \label{eq:onesampthresh}
C = 2^{25} \textrm{ samples per workunit } \times 14208 \textrm{ DMs } \times 2 \textrm{ signs } \times 2 \textrm{ from co-adds } = e^{28.3}.
\end{equation}

Since the standard deviation for the exponential is $\sigma = 1$, this means that by setting the threshold at $28. 3$  we would be looking for pulses that are $28.3 - 1 = 27.3 \sigma$ above the mean.

\begin{figure}[!h]
\includegraphics*[width=6.0in, viewport=0.2in 0.2in 6.4in 3.7in]{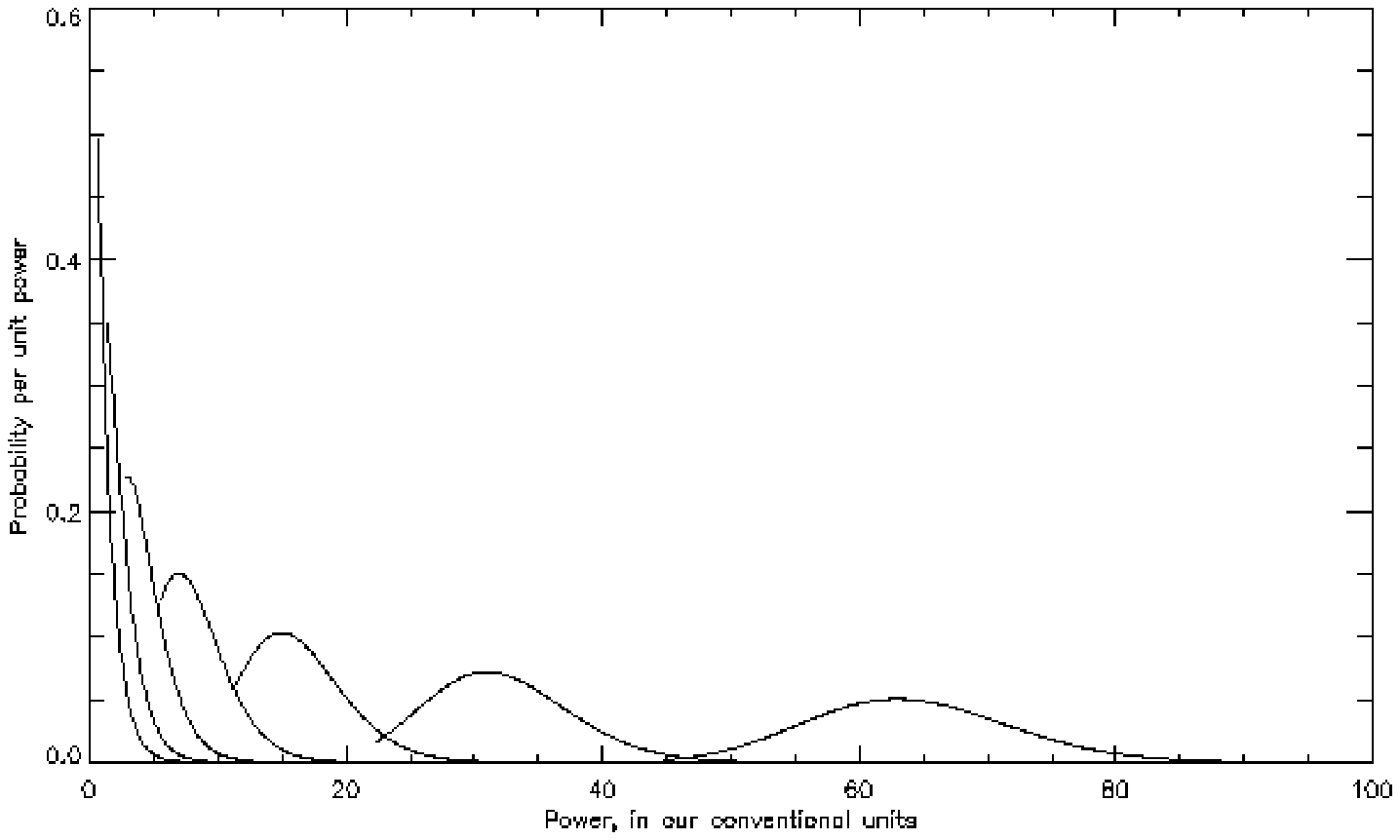}\caption[Gamma distributions]{Gamma distributions.  The x axis is integrated power (divided by $2$, as per our convention), and the y axis is probability per unit power.  The leftmost distribution, which is exponential,
belongs to $n = 1$.  The rest are $n = 2, 4, 8, 16, 32, 64$.  Notice that the
rightmost distribution is nearly a normal distribution.}
 \label{fig:gamma}\end{figure}

When we co-add $n = 2^\ell$ samples to make one co-added potential pulse, we are adding up that many exponential distributions.
The resulting power has a gamma distribution, with scale parameter $1$ and shape parameter $n$.  The pdf is

\begin{equation} \label{eq:gammapdf}
\frac{1}{\Gamma(n)} x^{n-1} e^{-x}.
\end{equation}

and the complementary cumulative distribution function is defined to be:

\begin{equation}
\int_x^\infty \frac{1}{\Gamma(n)} x^{n-1} e^{-x} dx = \frac{\Gamma(n, x)}{\Gamma(n)}.
\end{equation}

where $\Gamma(n, x)$ is the upper incomplete gamma function.  The first few pdfs are shown in Figure~\ref{fig:gamma}.

Then for each $n = 2^\ell$, we want to select a threshold, $H_n$, such that $\frac{1}{C} = \Gamma(n, H_n)/\Gamma(n)$.  We admit pulses of width $n$ only if they have power greater than or equal to $H_n$.  Since a pulse of power $H_n$ would occur by chance with probability $\frac{1}{C}$, and there are precisely $C$ potential pulses in a workunit, we expect to obtain $C \cdot \frac{1}{C} = 1$ false positive per workunit.

Because the probability distribution comes from a sum of identical exponential distributions, we can say that for co-add $n = 2^\ell$, the variance is $n$ times higher, and the standard deviation is $\sqrt{n}$ times higher, than for an exponential.  If we define $m$ to be the number of standard deviations of our threshold above the mean, then we are looking for pulses at $m_n = (H_n - n)/\sqrt{n}$.  A computation of $m_n$, using our actual thresholds $H_n$ (as determined by simulation, rather than theory) can be found in Table~\ref{tab:thresholds}.

\subsubsection{Expected discrepancies with the model}

A few differences from the model can be expected:

\begin{enumerate}

\item \textbf{Hydrogen line and filter shape}.  We assumed above that the input data is white noise.
In practice, this is not the case, because a portion of our band has higher
power due to the hyperfine hydrogen line.  The strength of this line can vary
depending on our RA and dec.  Then $\tilde{f}(k)$ no longer have
equal standard deviations.  This will cause some correlation between the deconvolved power
of adjacent samples, which will modify the pdf of the binned power, increasing the
variance.

To see this, consider the simplest, most extreme case: we imagine that the hydrogen line takes the form of a strong delta
function in the frequency domain of amplitude $A$ at frequency $k_0$, where $A$ is distributed randomly
according to a
Gaussian distribution with standard deviation $\sigma$ and mean $0$.  Then if $f_d$ is the dechirped
amplitude in the time domain, $f_d(t) = A e^{2 \pi i k_0 t / N}$.  (In other words, the dispersion is not relevant, since the hydrogen line has a single frequency, and we are momentarily assuming it overwhelms the noise.)  So $|f_d (t)|^2$ is exponential
with power $\sigma^2$, as discussed in Section~\ref{sec:singlepulsetheory}.  Now consider two nearby times $t_1 \sim t_2$ such that the phase of the hydrogen line does not change much between these samples.  We want to sum the amplitudes $f_d(t)$ at these nearby times in order to build a co-add.  Then $|f_d(t_1) + f_d(t_2)|$ is 
$A |e^{2 \pi i k_0 t_1 / N} + e^{2 \pi i k_0 t_2 / N}|$.  Since the phases are similar, this is roughly $2 A$, which
is Gaussian with standard deviation $2 \sigma$ and variance $4 \sigma^2$.  Whereas if we
summed samples without the hydrogen line, adding identical and independently distributed (iid) exponentials, the variances would simply add to give
$2 \sigma^2$.  So this model hydrogen line increases the variance.

In actuality, the effect of the hydrogen line is not so pronounced, but the idea is similar.
In the same way, the nonuniform shape of our low pass filters also causes the signal to differ from
white noise.

\item \textbf{Other disparities}

Even in the absence of the hydrogen line, tests reveal other differences between
the theoretical and actual distributions.  For high co-adds, the variance is
slightly less than expected.

It's easy to see that power per sample cannot be independently distributed, even
in the case of white noise.  This is because the total power over all samples must
be a constant; in our case, the constant is $32{,}768 = 2^{15}$, the total number of
samples in a FFT.  This would certainly result in a smaller variance, but
we have not established whether this effect suffices to explain the observed disparity.

\end{enumerate}

\subsubsection{Single pulse thresholds: Monte Carlo simulation}

To choose our thresholds for the single pulse search, we ran the client on $10$ ``noise'' workunits, which we had constructed to contain only white noise.  We kept track of the strongest pulses we found at each co-add.  The second largest pulse out of $10$ is roughly the $90$th
percentile, so we set our thresholds at that point. This method gives thresholds that are reasonable as long as we don't demand that we detect precisely equal numbers of pulses at each co-add.  The thresholds suggested by this simulation were within $1$\% to $5$\% of the theoretical values.

Because the ``noise'' workunits contained only white noise, with no Hydrogen line, some of the deviations described above would not be expected to occur.  For this reason, our simulation was an imperfect model.  However, our concern in this case was not to model the noise and Hydrogen line perfectly, but to obtain a rough estimate for the correct thresholds.  The Hydrogen line looks slightly different at different points on the sky, so a perfect simulation would be impossible in any event.  As discussed above, our thresholds are supplemented by our RFI and noise rejection algorithms, so we have some flexibility in setting thresholds.

\begin{table}[ht]
\caption{Pulse area thresholds $H_n$, in normalized units such that one sample has an expected power of $1$ unit, derived from the Monte Carlo simulation.  We denote the implied number of standard deviations above the mean by ``$m$''}
\centering 
\vspace{\baselineskip}
\begin{tabular}{c c c c} % centered columns (4 columns)
\hline\hline          
$\ell$ & $n$ & $H_n$ & $m$ \\ [0.5ex]  
\hline             
0 & 1 & 29.1 & 28.1 \\ 
1 & 2 & 31.6 & 20.9 \\ 
2 & 4 & 37.9 & 16.6\\ 
3 & 8 & 49.4 & 14.6 \\ 
4 & 16 & 61.3 & 11.3\\ 
5 & 32 & 87.0 & 9.7\\ 
6 & 64 & 128.9 & 8.1 \\ 
7 & 128 & 212.6 & 7.5\\ 
8 & 256 & 362.0 & 6.6\\ [1ex] 
\hline
\end{tabular}
\label{tab:thresholds} % is used to refer this table in the text
\end{table}

% Sensitivity comparsion

\subsection{Nominal sensitivity} \label{sensitivity}

\subsubsection{Scattering} \label{sec:sensscat}

In this section, we will discuss the nominal sensitivity of Astropulse and other surveys.  By this, we mean the minimum pulse area of an astrophysical pulse that can pass the thresholds described in Section~\ref{thresholds}.  Such a pulse might still be RFI or noise; we will discuss schemes for further rejecting RFI and noise in Section~\ref{sec:rfimitigation}.  (Nevertheless, we do not intend to claim a single pulse near threshold as a detection; see Section~\ref{sec:limitations}.)

We will first consider scattering.  Scattering impacts the sensitivity of a survey by limiting its resolution.  We suppose that an instantaneous pulse would be broadened by scattering to a width $\Delta t_\mathrm{sc}$.  Then we can estimate $\Delta t_\mathrm{sc}$ using the empirical formula given in \cite{lor05}:

\begin{equation} \label{eq:DMlimit}
\log \Delta t_\mathrm{sc, ms} = -6.46 + 0.154 (\log \DM) + 1.07 (\log \DM)^2 - 3.86 \log \nu_\mathrm{GHz}.
\end{equation}

However, this formula applies to sources in the Milky Way.  Astropulse spends a
substantial fraction of the time looking away from the plane of the Galaxy.  Some pulses may
originate far outside the Galaxy, in which case
$\Delta t_\mathrm{sc}$ should be much smaller, even for large DMs.  But the distribution
of the intergalactic medium (IGM) is not well understood.  \citet{lov07} find that
extragalactic radio sources at redshifts greater than $z = 2$ do not have
microarcsecond structure, suggesting that they are scatter broadened by turbulence
in the IGM.  This fact can be used to estimate the width $\Delta t_\mathrm{sc}$ of the broadened pulse.
A microarcsecond of angular broadening at $z = 2$ corresponds to a pulse
width of $\Delta t_\mathrm{sc} = \theta^2 d / c = 8 \us$ (using $d = 3.57$ Gpc).  According to
\cite{iok03}, this is at a DM of about $2000 \pccm$.  So perhaps we can assume that
at the (smaller) DMs of our experiment, pulses will have reasonably small widths.
For instance, for a source originating in or near the Galaxy, a DM of $830 \pccm$ would have a scattering width $400$ times
smaller than a DM of $2000 \pccm$.  Even if the scattering width were nearly $8 \us$,
Astropulse is still good at detecting such pulses.
(The threshold is just twice as high as for $1$ sample pulses.)

\subsubsection{Nominal sensitivity of Astropulse} \label{sec:sensastr}

To calculate Astropulse's nominal sensitivity as a pulse area in $\jyus$, we can 
follow the treatments in \citet{roh00} and \citet{van66}, which discuss the effect of
``clipping'' a noisy analog signal, changing it into a one-bit digital time series.

From these sources we conclude that if $F$ is the flux density and $N$ is the duration in samples of the minimal detectable pulse, then its pulse area is:

\begin{equation}
F \cdot (N t_\mathrm{sample}) = \frac{\pi}{2} \frac{T_0 \; t_\mathrm{sample} (H_N - N)}{G}. \label{eq:sensitivity}
\end{equation}

In this expression:

\begin{enumerate}
\item $H_N$ is the power threshold derived from the gamma distribution in Section~\ref{sec:singlepulsetheory}, and is dependent on $N$.
\item $T_0 \sim 30\; K$ is the system temperature.\footnote{\verb,http://www.naic.edu/alfa/gen_info/info_obs.shtml,} 
\item $G = 10$ K Jy$^{-1}$ is the telescope gain, roughly equal to $\frac{A}{k}$, where: 
\item $A = \frac{\lambda^2}{\Omega}$ is the effective area of the telescope and $k$ is Boltzmann's constant.  
\item $\lambda = 21$ cm is the wavelength of the signal.
\item $\Omega = 8.1 \cdot 10^{-7}$ is the beam's solid angle.
\end{enumerate}

So the resulting pulse area is $1.9\;(H_N - N) \jyus$.

\subsubsection{Nominal sensitivity comparison} \label{sec:senscomp}

While Astropulse detects a signal coherently, other undirected radio surveys use
 incoherent detection schemes.  Typically they use a filter bank, dividing the 
spectrum into $N$ sub-bands as described in Section~\ref{dedispersion}.  
\cite{den08} give the sensitivity formula as:

% Note: Phinney & Taylor seem to assume that some papers have an integration time tau,
% and are ignoring all parts of the pulse outside region tau, so that the
% sensitivity formula is prop to t / sqrt(tau) rather than t / sqrt(t)

\begin{eqnarray} \label{eq:denevatemp}
S_\mathrm{min} = (\frac{W}{W_i}) \frac{m S_\mathrm{sys}}{\sqrt{N_\mathrm{pol} B W}}, \\ 
\label{eq:effwidth}
W = (W_i^2 + \Delta t^2_\mathrm{DM, ch} + \Delta t^2_\mathrm{DM, err} + \Delta t^2_\mathrm{sc})^{1/2}.
\end{eqnarray}

\begin{list}{}{}

\item $S_\mathrm{sys}$ is the system-equivalent flux density.

\item $m$ is the desired number of standard deviations for the detection threshold,

\item $W$ is the effective width of the pulse, including broadening due to dispersion and scattering.

\item $W_i$ is the intrinsic width of the pulse, prior to broadening.

\item $\Delta t_\mathrm{DM, ch}$ is the dispersion within one channel.

\item $\Delta t_\mathrm{DM, err}$ is the error caused by looking at the wrong dispersion measure.  We have a DM error of $\frac{1}{2}$ the DM step.  (The time error $\Delta t_\mathrm{DM, err}$ depends on the bandwidth as well as the DM error.)

\item $\Delta t_\mathrm{sc} \propto f^{-3.86}$ is the error caused by scattering broadening, where $f$ is the pulse's frequency.

\end{list}

Equation~\ref{eq:denevatemp} assumes that the channel bandwidth is not so narrow that we are sampling 
beyond the Nyquist rate.  If the channel bandwidth were that narrow, there 
would be another contribution to the effective width; but all surveys are
careful not to sample beyond the Nyquist rate.

If one cares about the pulse area in $\jyus$ rather 
than the instantaneous flux density, a simpler expression will suffice.  If one has observed at a single polarization, the minimum detectable pulse area for that polarization will be
$A = \frac{m T_0 t}{2 G \sqrt{B t}} = m \frac{T_0}{2 G} \sqrt{t / B}$, and the pulse area of both polarizations together
will be $A = m \frac{T_0}{G} \sqrt{t / B}$.  If both polarizations are observed together and their power is summed, then the minimum detectable pulse area of the two polarizations together is smaller by a factor $\sqrt{2}$, assuming an unpolarized signal.
% One must also compute $t$ from Equation~\ref{eq:effwidth}.

As discussed in Section~\ref{sec:sensscat}, the scattering error is likely to become less important when we are observing pulses that originate far outside the Galaxy.  So, with an understanding that different conditions would apply when observing pulses that originate in or near the Galaxy, we will ignore the scattering error in Table~\ref{tab:survey} and set $t = t_\mathrm{sample}$.  In Table~\ref{tab:survey}, we list characteristics of Astropulse and other surveys and the following conventions are observed:

\begin{itemize}

\item $m$: number of standard deviations for threshold.  For any survey that uses incoherent dedispersion, this refers to the standard deviation of a Gaussian distribution, so $m = 6$ suffices to rule out all but $1$ in $10^9$ spurious detections due to noise.  However, for Astropulse, the distribution is a chi-square, as discussed in Section~\ref{thresholds}.  In the worst-case scenario -- a one-sample pulse -- this chi-square is equivalent to an exponential, and $m = 23$ is required to rule out all but $1$ in $10^9$ spurious detections due to noise.  In fact, we selected the value $m = 30$ for Astropulse in the case of one-sample pulses.  For all surveys, the optimal value of $m$ is not determined only by the statistical distribution, but also by RFI and noise rejection.  For Astropulse, the statistics are discussed at length in Section~\ref{sec:singlepulsetheory}, and RFI and noise rejection are discussed in Section~\ref{sec:rfimitigation}.  In Table~\ref{tab:survey}, we have listed the values of $m$ reported by each survey in available publications.  If we could not determine the survey's $m$-value, we assumed $m=6$.  We are assuming that astrophysical pulses which surpassed each survey's stated threshold would be detected by the survey and could be distinguished from noise and RFI.

\item $t_\mathrm{sample}$: the time resolution of the survey.

\item $t$: the minimum effective duration of a pulse after dedispersion.

\item beam $\Omega$: the telescope beam width, in steradians.

\item beams: number of simultaneous beams.

\item $t_{\mathrm{obs}}$: observation time per beam, in hours.

\item $N_\mathrm{pol}$: Astropulse detects a pulse using data from a single polarization; many surveys combine two polarizations.

\item sens: the minimum detectable pulse area.  For each survey, the listed sensitivities apply for pulses which are unresolved by that survey.   Pulses of duration $0.4 \us$ are unresolved by all surveys, including Astropulse.  For pulses of greater duration, Astropulse's sensitivity will degrade.  Table~\ref{tab:thresholds}, discussed earlier in Section~\ref{sec:singlepulsetheory}, depicts the relative increase in Astropulse's minimum detectable pulse area for wider pulses.  Other surveys' sensitivities do not degrade until their time resolution is reached.

\end{itemize}

In order to compare the surveys using a concrete example, we will also provide information about each survey's ability to detect evaporating primordial black holes, under specific assumptions: that $M = 10^8$ kg of the black hole's mass is transformed into a radio signal of bandwidth $~1$ GHz, and that scattering broadening does not substantially interfere with detection.  These assumptions are not intended to be representative of all models of evaporating black holes, and black holes themselves are not the only potential source of microsecond pulses.

\begin{itemize}
\item $d_\mathrm{max}$: the minimum distance from which an exploding
$M = 10^8$ kg black hole would be visible, in kpc.  It's calculated using

\begin{equation} \label{eq:sens_energy_dist}
U_\mathrm{min} = \mathrm{ energy } / (\mathrm{area } \cdot \mathrm{ bandwidth})
 = (M c^2) / (4 \pi d_\mathrm{max}^2 \cdot 1 \mathrm{GHz}),
\end{equation}

where $U_\mathrm{min}$ is the
pulse area of the minimum detectable signal in $\jyus$.

\item rate: the minimum rate of black hole explosions under which such a black
hole would be detectable, $V^{-1} t_\mathrm{obs}^{-1}$.  Here,
$V = (4 \pi / 3) d_\mathrm{max}^3 n_\mathrm{beams} \frac{\Omega}{4 \pi} =
\frac{1}{3} \Omega d_\mathrm{max}^3 n_\mathrm{beams}$ is the volume of space observed at any one time.

\end{itemize}

We conclude that Astropulse's minimum detectable rate (in black holes explosions pc$^{-3}$ yr$^{-1}$) is 
comparable to that of other surveys, but not superior.  Astropulse's rate is similar to
\citet{lor07} and \citet{den08}.  
Our sensitivity to unresolved pulses, in $\jyus$, is superior to all
other surveys listed except for the Arecibo multibeam survey of \citet{den08}.  This sensitivity
comes largely from our microsecond time resolution and high gain.  Our observation time is also superior.  
Astropulse does have substantial disadvantages, including a limited bandwidth and narrow
($\Omega = 8.1 \cdot 10^{-7}$) beams.

\begin{center}

\clearpage

%% We can't use deluxetables here
%% because some guy decided that deluxetables
%% should have no positioning options
%% therefore I had to rewrite my deluxetables
%% as ordinary tables here.

\addtocounter{footnote}{1}

%%%%% TABLE NUMBER 1 %%%%%%%
\begin{sidewaystable}[!h]
\caption[Survey parameters]{Survey parameters.  Parentheses around a value indicate that we assume this value because we could not deduce one from the original paper.}
\vspace{10mm}
\begin{minipage}{\textwidth}
\renewcommand{\thefootnote}{\emph{\alph{footnote}}}
\begin{center}
\begin{tabular*}{0.9\textwidth}{lllclcccccc}
\hline
\# & author & telescope & year & dedisp & ref & 
$\nu_0$ (MHz) & $m$ &
$T_0$ (K) & $t_\mathrm{sample} (\mu s)$ & $t (\mu s)$ \\
\hline
1 & O'Sullivan et al. & Dwingeloo & 1978 & incoh & \footnotemark[1] &
$5000$ & $(6)$ & $65$ & $2$ & $2700$ \\
2 & Phinney \& Taylor & Arecibo & 1979 & incoh & \footnotemark[2] &
$430$ & $6$ & $175$ & $1.7 \cdot 10^4$ & $1.7 \cdot 10^4$ \\
3 & Amy et al. & MOST & 1989 & incoh & \footnotemark[3] & 
$843$ & $(6)$ & - & $1$ & $1.7 \cdot 10^4$ \\
4 & Katz \& Hewitt & STARE & 2003 & incoh & \footnotemark[4] &
$611$ & $5$ & $150$ & $125000$ & $125000$ \\
5 & McLaughlin et al. & Parkes & 2006 & incoh & \footnotemark[5], \footnotemark[6] &
$1400$ & $5$ & $21$ & $250$ & $250$ \\
6 & Lorimer \& Bailes & Parkes & 2007 & incoh & \footnotemark[7] &
$1400$ & $(6)$ & $21$ & $1000$ & $1000$  \\
7 & Deneva et al. & Arecibo & 2008 & incoh & \footnotemark[8] &
$1440$ & $5$ & $30$ & $64$ & $64$ \\
8 & Von Korff et al. & Arecibo & 2009 & coher & - &
$1420$ & $30$ & $30$ & $0.4$ & $0.4$ \\
\end{tabular*}

\vspace{\baselineskip}
\begin{tabular*}{0.9\textwidth}{llccccccccc}
\hline\hline
\# & $\Delta \nu$ (MHz) & G (K Jy$^{-1}$) &
beam $\Omega$ & beams & $ t_\mathrm{obs}$ (h) &
$N_\mathrm{pol}$ & sens ($\jyus$) &
$d_\mathrm{max}$ (kpc) & rate (\pcyr) \\
\hline
1 & $100$ & $0.1$ & $6.6 \cdot 10^{-6}$ & $1$ & $46$ &
$1$ & $2 \cdot 10^4$ & $61$ & $3.8 \cdot 10^{-7}$ \\
2 & $16$ & $27$ & $6.6 \cdot 10^{-6}$ & $1$ & $292$ & 
$1$ & $1300$ & $240$ & $9.4 \cdot 10^{-10}$ \\
3 & $3$ & - & $3.6 \cdot 10^{-8}$ & $32$ & $4000$ &
$1$ & $1.6 \cdot 10^5$ \footnotemark[9] & $22$ & $5.6 \cdot 10^{-7}$ \\
4 & $4$ & $6.1 \cdot 10^{-5}$ & $1.4$ & $1$ & $13000$ &
$2$ & $1.5 \cdot 10^9$ & $0.22$ & $1.3 \cdot 10^{-7}$ \\
5 & $288$ & $0.7$ & $1.3 \cdot 10^{-5}$ & $13$ & $1600$ & 
$2$ & $99$ & $870$ & $1.5 \cdot 10^{-13}$ \\
6 & $288$ & $0.7$ & $1.3 \cdot 10^{-5}$ & $13$ & $480$ &
$2$ & $240$ & $560$ & $1.8 \cdot 10^{-12}$\\
7 & $100$ & $10$ & $8.1 \cdot 10^{-7}$ & $7$ & $420$  &
$2$ & $8.5$ & $3000$ & $4.2 \cdot 10^{-13}$ \\
8 & $2.5$ & $10$ & $8.1 \cdot 10^{-7}$ & $7$ & $1540$ &
$1$ & $55$ & $1200$ & $1.9 \cdot 10^{-12}$\\
\hline
\end{tabular*}
\end{center}
\footnotetext[1]{\cite{osu78}}
\footnotetext[2]{\cite{phi79}}
\footnotetext[3]{\cite{amy89}}
\footnotetext[4]{\cite{kat03}}
\footnotetext[5]{\cite{mcl06}}
\footnotetext[6]{\cite{man01}}
\footnotetext[7]{\cite{lor07}}
\footnotetext[8]{\cite{den08}}
\footnotetext[9]{ MOST has $1$ mJy of noise in each beam after $12$ hours, http://www.physics.usyd.edu.au/sifa/Main/MOST}
\end{minipage}
\label{tab:survey}
\end{sidewaystable}
% \vspace{10mm}
\clearpage

%% O'Sullivan:
%% The paper says "the sensitivity was ~0.1 K Jy^{-1}".  I had written down 0.4 K Jy^{-1}.
%% Why did I write that?
%%
%% Amy:
%% uncertain whether Amy really has 32 beams and 4000 hours, or 32 and 4000/32 hours, or what.
%% where does < 7 gain come from?  info on Molonglo?
%% Katz:  why 40000 hours?  18 months x 30 days x 24 hours is more like 13000.
%% paper says ``18 months of round-the-clock observing''
% 4 & Katz \& Hewitt & $4$ & $1$ & ? & 40000? \\
%% McLaughlin: 35000 hours is 4 years, and comes from the RRATs paper,
%% but before I had 20000 / 13 = 1500 hours.  Where did I get that from?
%% Answer: we examine 10 degrees by 150 degrees, 35 minutes observation of each location
%% (35 min figure found in both lor07 and man01).  Beam is 1.3 * 10^-5 ster.
%% then you can derive the time duration of the survey as 1600 hours.
%% see McLaughlin bottom of page 818 (the second page of the paper)
%% under Figure 1
%% Deneva 3000 hours is calculated in the same way.

% \vspace{10mm}

%% d: sqrt(10^8 * 9 * 10^16 / (4 * pi * sens * 10^-32 * 10^9)) / (3 * 10^16)
%% rate: 1/(1/3*(d * 1000)^3 * Omega * nbeams * tobs / 24 / 365)

%% When http://www.physics.usyd.edu.au/sifa/Main/MOST says "polarization: RH circular"
%% does that imply one polarization only?
%% No strong evidence for Npol in (1), (2)

\end{center}

% How we rejected the radio frequency interference
%\subsection{RFI rejection} \label{rfirejection}
%\input{rfirejection}

% BOINC
% \subsection{BOINC} \label{boinc}
% \input{boinc}

% Testing and verification (birdie detection, known pulsars, hydrogen)
% \subsection{Testing and Verification} \label{verification}
% \input{verification}

\section{Distributed computing : the BOINC platform} \label{boinc}

\begin{figure}[!h]
\includegraphics[width = 6.0in]{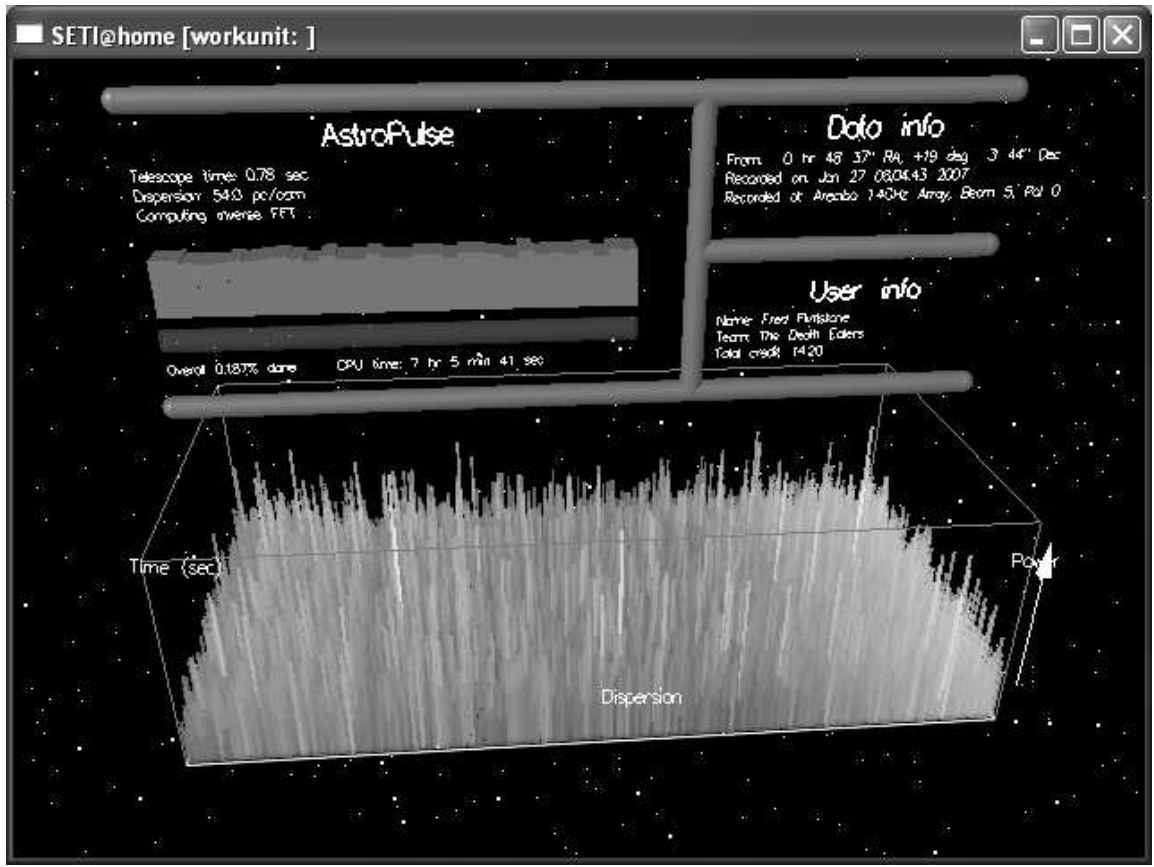}
\caption[The Astropulse screen saver]{The Astropulse screen saver.}
 \label{fig:AstropulseScreenSaver}
\end{figure}

Astropulse runs on the BOINC platform \citep{and04}, an acronym for ``Berkeley Open Infrastructure 
for Network Computing.''  BOINC is a set of programs that organizes volunteers' home
computers to perform scientific calculations.  In a typical BOINC project, a researcher
has a computing problem that can run in parallel, that is, on several machines at once.
Perhaps the problem involves searching a physical space (for Astropulse, this space is the sky), and performing the
same computation on each point in that space (for Astropulse, this computation is dedispersion.)  The first BOINC project, SETI@home, searched the sky for
narrowband transmissions.  The space could also be a parameter space, for instance a space of potential
climate models (climateprediction.net) or protein configurations (Rosetta@home).  The visible manifestation of a BOINC project is an informative screen saver.  Figure~\ref{fig:AstropulseScreenSaver} shows the Astropulse screen saver.

Although the volunteers are providing their computers for free, the bandwidth and storage space required to distribute data to the volunteers is not free.  A project is suitable for BOINC only if it is computation-intensive.  That is, the monetary cost to perform the computation must be greater than the monetary cost of distributing the data.  Coherent dedispersion satisfies this requirement, because we must perform FFTs at many DMs.

The researcher for a BOINC project need not be affiliated with UC Berkeley, or with the BOINC development
 team at Berkeley (although we happen to be so affiliated).  BOINC is open source, and can be 
downloaded, compiled, and operated by anyone with sufficient technical skills; about $50$ projects
currently exist outside Berkeley.  

Likewise, volunteers need not have any particular technical knowledge.  They just have to navigate
to the BOINC web page with their web browser, and follow the instructions to download
the Astropulse ``client'' program.  Astropulse has
access to around 500,000 volunteers, each of whose machines might have $2$ GFLOPs of processing
power, and be on $1/3$ of the time, for a total of $300$ TFLOPs -- as much as the world's fastest
general purpose supercomputer in 2007, IBM's Blue Gene / L\footnote{\verb,http://www.top500.org/list/2007/06/100,}.  Since that time, the processing power of the fastest supercomputer has increased to $8000$ TFLOPs or more\footnote{\verb,http://www.top500.org/list/2011/06/100,}.

\section{RFI and noise mitigation} \label{sec:rfimitigation}

The current section describes the methods by which we have classified all detected pulses, deciding whether they might correspond to RFI or noise.  This section is largely about RFI rejection -- rejecting spurious events while throwing away as few real events as possible.  However, it is equally important to reject noise events, and in Section~\ref{sec:multipols}, we compute the odds that a purely noise event will pass our tests.  We will begin by discussing a number of RFI and noise mitigation methods.  We will then define a figure of merit statistic for each method, and use Monte Carlo simulations to argue that astrophysical signals will not be excluded by our RFI and noise mitigation methods.  Finally, we will mention our successful detection of giant pulses from the Crab pulsar, providing further evidence that astrophysical signals will not be excluded; and we will discuss limitations of this study.

The dominant RFI sources at Arecibo Observatory are nearby radars, which emit one of (at least) $6$ repeating patterns.  These include the Federal Aviation Administration (FAA) radar, used for air traffic control around Puerto Rico; the aerostat radar, used for drug interdiction; and others.

\subsection{RFI and noise mitigation methods} \label{sec:rfimethods}

We rejected RFI and noise using several methods.  We will first discuss methods implemented prior to pulse detection.  Most of these methods (\ref{sec:hardwareblanker}, \ref{sec:softwareblanker}, and \ref{sec:clientblanker}) blanked segments of the data which were likely to be contaminated by radio frequency interference.  We will then describe a number of post-detection techniques.  Using these techniques, we characterized pulses as probable RFI by examining their dispersion measures, polarizations, shapes, and other properties, and by examining other pulses detected at nearby times.

\subsubsection{Arecibo's high pass filter}

Arecibo can turn on a high pass filter in the receiver that will reject the FAA radar's band from the data.  However, Astropulse operates commensally, and our partner surveys usually require that this filter is turned off.

\subsubsection{Hardware blanker} \label{sec:hardwareblanker}

Arecibo Observatory provides us with a blanking signal (which we will call the ``hardware'' blanking signal), a single bit 
which is turned on when the FAA radar is transmitting a pulse.  The ``hardware''
blanker has two components:

\begin{enumerate}
\item The hardware component at Arecibo, which adds the blanking bit to our tape files.
\item A software component, activated at a later point in our data processing pipeline, which detects the bit and blanks the appropriate data.  (This software component still counts as a part of our ``hardware'' blanker.)
\end{enumerate}

It is critical that we blank the data using noise that has the same frequency profile
as the clean data.  If we instead blank the data using white noise (bits set randomly to
$1$ and $0$) artificial signals will be introduced, since the white noise does not match the rest of the data.
Therefore, we need to blank the data by replacing it with noise whose frequency envelope matches that of our data recorder's $2.5$ MHz bandpass filter.
% To do this, we sample the unblanked data, take its Fourier transform, and multiply by our filter shape to construct
% an appropriate stream of shaped noise.

Unfortunately, the hardware blanker is imperfect.  First, we believe it doesn't mark every
FAA radar pulse.  Sometimes the radar's phase changes, and it takes some time for the hardware
blanker to catch up.  At other times, a single radar pulse may arrive that is out of sync with
the other pulses.  Second, the hardware blanker only searches for the FAA radar, not
for other radar.
So we have written our own software blanker, which processes the data downstream from the hardware
blanker.  The software blanker handles both the FAA radar and the aerostat radar.

\subsubsection{Software blanker} \label{sec:softwareblanker}

The software blanker is an algorithm that runs on a computer in Space Sciences Lab.  It examines the data for the repeating patterns that signify radar.  It looks specifically for the FAA and aerostat radar.

To find either radar, the blanker looks at samples in groups of $10$.  A radar pulse would consist of samples where the bits are predominantly $1$ or predominantly $0$.  (That's 280 bits, counting all 28 bits for each sample, 2 polarizations, 7 beams, and both real and imaginary bits.)  At maximum strength, the radar will produce long strings of bits that are all set to $1$, regardless of whether they represent real or imaginary data.  At lesser strengths, the radar produces less skewed sets of samples, with a ``ring down'' oscillation between 1 and 0 bits.  Nevertheless, these sets of samples at lesser strengths can provide an important indication of radar.  

The blanker folds the data over $25$ seconds at the known radar periods, which are $35{,}262$ samples for the FAA radar, and $57{,}571$ samples for the aerostat radar.  (For each radar source, the intervals between pulses vary according to a set sequence.  So the stated FAA period actually contains $5$ pulses, and the aerostat period contains $7$ pulses.)  Actually, we fold at around $200$ trial periods, each varying slightly from the average radar period.  This is necessary because the radar's period can drift slightly.  We threshold the resulting amplitudes at 25\% above the mean.  

If a radar signal has been detected, we blank data at regular intervals over the $25$ seconds, accounting for the varying interpulse periods.  Typically, the radar shape is square: a total radar silence, followed by radar of duration $800$ samples, followed by more silence.  However, we have observed that individual radar pulses may be smeared by up to $50$ samples on either side of the region of duration $800$.  Therefore, we blank $100$ samples on either side, for safety.  In total, we typically blank $1{,}000$ samples for each radar pulse.

Most of our data contains FAA radar signals, and relatively little has aerostat signals.  Overall, we blank about $14\%$ of our data during this stage.  The software blanker is very effective in comparison with the hardware blanker; more than $99\%$ of pulses that are detected by the hardware blanker are also detected by the software blanker, whereas the reverse is not true.

\subsubsection{Client blanker} \label{sec:clientblanker}

The software blanker is effective at removing FAA and aerostat radar patterns from our data, but other types of RFI exist which do not have these periods.  In order to detect and remove these other signals, we implement a blanker in the Astropulse client.   Like the software blanker, the client blanker searches for RFI that is strong enough to saturate our electronics, producing a long string of identical samples.  We have found that this string of identical samples not only indicates RFI at that particular instant, but also warns that other nearby data in the same workunit may be contaminated by RFI.  Although a string of identical samples is not itself a dispersed pulse, it can signify that dispersed RFI may be present in the nearby data.  Astropulse could record this dispersed RFI as a candidate pulse in our database.  Therefore, we blank all data within $400{,}000$ samples ($0.16$ s) of the detected event.  This figure, $400{,}000$ samples, was determined empirically, by examining many workunits showing strings of identical samples.  We found that these strings were often accompanied by dispersed pulses within a distance of $400{,}000$ samples.  We do not believe that these dispersed pulses represented true astrophysical transients, because they were very common, and because, at least for our system, the strings of identical samples are a known feature of RFI.

The client blanker differs from the software blanker in that we consider individual RFI events, rather than folding several events together.  This enables detection of RFI with unknown periods.  The client blanker proceeds by performing a Fourier transform of a segment of the data, and examining the power in the central bin (the DC component.)

\subsubsection{Fraction blanked restriction}

This and all following methods occurred in the post-detection phase; that is, we stored pulses in our database prior to determining whether they passed or failed this test.  For the current method, we consider each workunit and record the fraction of the data that we blanked using the
client blanker.  
We remove workunits entirely if too much RFI was present, since we have observed that the presence of too much RFI in one region of the workunit may indicate some amount of RFI in other regions.

\subsubsection{DM repetition} \label{sec:dmrepetition}

If we see a signal at the same DM repeatedly over a short period of time and at different parts of the sky,  we conclude that 
it came from a terrestrial source and reject these signals as RFI.  There is no reason that the same DM should have been observed
from several different directions in quick succession unless the source was terrestrial.  This test, like the multi-beam and multi-polarization tests described below, is implemented by a program that examines our database of detected candidate pulses, putting them in time order and searching for the appropriate pattern.

Note that it is possible for a single astrophysical source to produce multiple candidate pulses (Section~\ref{sec:crab}.)  One might be concerned that these pulses would fail the DM repetition test.  However, the DM repetition test does not trigger unless pulses having the same DM are detected more than $25$ arcminutes apart on the sky (wider than the distance between any two beams).  Therefore, the test cannot be triggered by multiple candidate pulses closely spaced in time.

\subsubsection{Multiple simultaneous beams}

This method, which makes use of the ALFA receiver's multiple beams, proved to be less useful than we had hoped.  We did not make use of it for the calculations discussed in this paper, but nevertheless we describe the method here.  The method relies on the fact that our beams are separated by several arcminutes, so an astrophysical source (or any relatively
weak source) should not appear
in multiple beams simultaneously.  However, a very strong source, for instance a terrestrial
source, might appear in the beams' sidelobes.  The source's radio waves might arrive at the
telescope by scattering from nearby terrain, or by bouncing off the
telescope support structure.  In this case, the waves might appear in multiple beams.
Since the telescope never points at or below the horizon, such sources would
appear only in the sidelobes and never in the main lobe.  

Therefore,
we could rule out some RFI by ignoring pulses that appear in multiple beams simultaneously.
Unfortunately, experiments show that a few real, astrophysical signals appear in multiple
beams simultaneously -- namely, the detected Crab pulses.  This may happen because the main lobes intersect slightly, albeit
at greatly diminished sensitivity, or because strong astrophysical signals could be detected
in sidelobes.  So such a method is imperfect at best.

\subsubsection{Two simultaneous polarizations} \label{sec:multipols}

A signal from an unpolarized astrophysical source will appear in both polarizations simultaneously
(unless it is only marginally detectable.)
RFI might also behave this way, but noise will not.  Therefore, we can reject a great deal of
noise by requiring detections in two simultaneous polarizations.  Unfortunately, highly polarized astrophysical signals may also be rejected, especially if the signal's axis of polarization lines up with the telescope's axis of polarization.  This drawback is balanced by the extraordinary efficacy of the polarization test as a noise rejection technique.

We argued in Section~\ref{thresholds} that only $e^{-28.3}$ of all noise pulses would pass through the thresholds we have designed, and that our goal should be to pass at most $e^{-43.1}$.  We ask next whether we can reject the required amount of noise.  Under the assumption that noise is uncorrelated between the two polarizations, we will show that we can reject all but $e^{-41.7}$ of noise pulses via techniques used in the current analysis.  This level of noise rejection would result in $4$ noise pulses passing our tests, making it impossible for us to claim a detection of a single astrophysical pulse at or near threshold.  A discussion of this limitation, including possible plans for working around it, are given in Section~\ref{sec:limitations}.

For a pulse to pass the multi-polarization test, it must have a partner in the corresponding data stream (from the other polarization).  To allow for the possibility that pulses may be detected at somewhat unpredictable times, we only require the two partners to be located in the data stream within $3$ times the pulse width (duration) of the larger pulse.  On average, we expect a single noise pulse in each data stream (see Section~\ref{thresholds}), and we assume for the purpose of this estimation that there is precisely one in each.  We take the noise pulse of smaller width to be at a given location, and we compute the probability that the wider noise pulse is near it (but in the other data stream.)  We will say that pulses have widths $w_S$ (small) and $w_L$ (large) with $w_S < w_L$.  Then the larger pulse can be at about $6$ locations if there is to be a coincidence; three upstream of the first pulse and three downstream of it.  The total number of possible locations in the data stream for the larger pulse is $N / w_L$, where $N$ is the number of samples in a workunit and $w_L$ is the number of samples for the pulse's width.  This means that the probability for the larger pulse to be near the location of the smaller pulse is $6 w_L / N$.  Furthermore, we note that the probability for a noise pulse to have width $w$ (where $w$ must be a power of two between $1$ and $512$) is approximately $\frac{1}{2 w}$.  Therefore, the probability of a coincidence in a given pair of workunits, assuming a single noise pulse in each, is found by summing over the $100$ possible pairs of widths ($w_i, w_j$).
We must compute $\sum_{i,j}$ (Probability that first width is $w_i$) $\times$ (Probability that second width is $w_j$) $\times$ (Probability that this pair of widths leads to a coincidence).  $w_S$ will as usual denote the smaller of the two and $w_L$ the larger.

\begin{eqnarray} 
& & \sum_{i,j} \frac{1}{2 w_i} \times \frac{1}{2 w_j} \times \frac{6 w_L}{N}  \\
& = & \sum_{i,j} \frac{6}{4 N w_S} \\
& = & \frac{1.5}{N} \times (\frac{19}{1} + \frac{17}{2} + \frac{15}{4} + \ldots + \frac{1}{512})
\end{eqnarray}

Where the last line is derived from noticing that (for instance) $1$ is the smallest number in $19$ of the $10 \times 10 = 100$ pairs, $2$ is the smallest number in $17$ pairs, and so on.  This sum is equal to $51 / N$.  Our workunits contain $N = 2^{25}$ samples, resulting in a coincidence probability of $0.00000152 = e^{-13.4}$.
Therefore, we will pass $e^{-28.3-13.4} = e^{-41.7}$ of all noise pulses, and $4$ pulses due to noise would be expected to pass our tests.

\subsubsection{Frequency profile}

We are looking for broadband pulses with a short intrinsic timescale.  Thus,
the pulse should have roughly the same mean power at all frequencies.  We perform a 
chi-square test to determine whether the mean power is the same everywhere.
However, the chi-square distribution is not a perfect description of the power vs. frequency distribution unless the power has a Gaussian distribution at each frequency.  In fact, the power should have an exponential distribution, not Gaussian.

The frequency profile test calculates a ``log\_prob'' statistic, which is the natural log of the estimated probability that this frequency profile would occur by chance.  However, if the chi square is inaccurate, so is the log\_prob.
Nevertheless, the log\_prob should decrease (and is negative) as the power becomes concentrated at particular frequencies.  Although the log\_prob has uncertain meaning in the absolute sense, its relative value is meaningful.

\subsection{Figure of merit}

We can assign a figure of merit to each RFI rejection algorithm, or to all algorithms together.  The figure of merit is defined as: 

(\% of astrophysical pulses passing) / (\% of all candidate pulses passing).  

% Another way of writing this is:

% (\% of pulses that are astrophysical after the algorithm is applied) / (\% of pulses that are astrophysical before the algorithm is applied). 

A candidate pulse is a detection that is above threshold and is placed in the Astropulse database.  It may or may not correspond to a signal from an astrophysical source.  (To be clear, the current section considers three distinct types of pulses: candidate pulses that are astrophysical, candidate pulses that are not astrophysical, and simulated ``astrophysical'' pulses that are not candidate pulses and do not come from the Astropulse database.)  

The purpose of an RFI rejection algorithm is to throw out spurious candidate pulses, while hopefully preserving most of the astrophysical candidate pulses.  
If the figure of merit is equal to $1$, the algorithm does not change the percent of candidate pulses that are astrophysical.  In other words, we could have achieved the same result by throwing out a random collection of our candidate pulses.  Therefore, an algorithm cannot be useful unless its figure or merit is greater than $1$.  

This definition of the figure of merit is not the only one imaginable.  For example, suppose we have $1{,}000$ candidate pulses, of which $100$ are astrophysical, and two algorithms.  The first algorithm cuts the list down to $10$ pulses, of which $9$ are astrophysical.  It has a figure of merit equal to $9$.  The second algorithm instead cuts the list down to $100$ pulses, of which $80$ are astrophysical.  It has a figure of merit equal to $8$.  Then one might prefer the latter algorithm, on the grounds that it yields more data to work with (even though a smaller percent of that data is good).

Nevertheless, this figure of merit is reasonable, and we calculate its value for each of our RFI rejection algorithms in the following sections.  In our calculations, we will assume that the vast majority of candidate pulses in the Astropulse database are due to noise, since if this large number of transient radio signals were of astrophysical origin, the phenomenon would have been reported by other surveys.  

Our method for estimating a figure of merit requires us to simulate ``astrophysical'' pulses in a Monte Carlo simulation.  Therefore, our computation makes the assumption that the simulated astrophysical pulses are similar enough to real (candidate) astrophysical pulses that the two types of pulses have the same passing fraction (and therefore would lead to the same figure of merit.)  One way in which simulated pulses differ from real ones is that real astrophysical pulses can generate multiple candidates in the Astropulse database, while simulated ``astrophysical'' pulses cannot.  However, this would not necessarily lead to a difference in the passing fractions for simulated and real astrophysical pulses.  The fact that there are multiple candidates does not, in and of itself, affect the passing fraction, as long as an individual simulated pulse is a good simulation of each individual candidate.  On the other hand, if there were interaction between the multiple candidates within our RFI and noise mitigation tests, such that each candidate's passing probability depends on when we detect the other candidates, then candidates could not be simulated as individual pulses.  However, only one of our tests compares groups of related pulses -- namely, the DM repetition test -- and this test cannot be triggered by multiple candidate pulses closely spaced in time.

In the following sections, we will compute a figure of merit for the post-processing methods.  We cannot compute a figure of merit for pre-processing methods, since we cannot count the number of candidate pulses (such as RFI) that we would have detected in absence of those methods.

\subsubsection{Fraction blanked restriction: figure of merit}

Empirically, it turns out that we can obtain the best figure of merit by passing only those workunits for which the fraction blanked (by the client blanker) is $< 20$\%.   Note that the client blanker has already removed a portion of each workunit.  Here, we do not consider the figure of merit resulting from the operation of the client blanker itself.  Rather, we are throwing out workunits for which a large portion has already been removed.
As of December 2009, the figure of merit statistics are as given in Table~\ref{tab:fom}.  Instead of simulating astrophysical pulses, we have counted the space available for such pulses in all workunits.  Only unblanked space may contain astrophysical pulses (or any pulses that originate outside the telescope), and we are assuming that the likelihood for an astrophysical pulse to appear in a workunit is proportional to the amount of unblanked space in that workunit.

%Space, in workunit-lengths, available for astrophysical pulses:

%x = total space for pulses = total of (1 - fraction\_blanked) = 2457187

%y = total space for pulses with fraction\_blanked $< 0.2$ = 1368632

%z = \% of astrophysical pulses passing = y / x = 0.556991
%
%Pulses in workunits analyzed so far for rfi:
%
%x2 = total number of pulses = 256085
%
%y2 = total number of pulses in workunits with fraction\_blanked $< 0.2$ = 122627
%
%z2 = \% of all pulses passing = y2 / x2 = 0.478852
%
%figure of merit = z / z2 = 1.16318

\subsubsection{DM repetition: figure of merit} \label{sec:dmrepfom}

To simulate the fraction of astrophysical pulses that would be accepted by the DM repetition algorithm, we performed a Monte Carlo study, generating a list of 37,572 pulses at random times.  We made use of this list of pulses to analyze the DM repetition, multiple beams, and simultaneous polarizations tests.  The random times were determined by considering the start times of actual workunits, then selecting a random time within that workunit.  Other relevant parameters, such as dispersion measure, beam, and polarization, were assigned a uniformly distributed random value within the allowed range.  (Pulse area was not relevant for this list of test pulses.)

Using the random dispersion measures, we counted the number of detected pulses with the same dispersion measure preceding and following the test pulses.  The test pulses were accepted or rejected using the same criteria as the DM repetition RFI rejection method.

Monte Carlo statistics for simulated astrophysical pulses, after $3$ passes through $12,524$ workunits, generating $37,572$ test pulses, are listed in Table~\ref{tab:fom}.

%total pulses = 37572
%
%passing DM repetition test = 35994
%
%fraction passing = 0.958001
%
%Pulses in workunits analyzed so far for rfi:
%
%total pulses = 204994
%
%passing DM repetition test = 114795
%
%fraction passing = 0.559992
%
%figure of merit = 0.958001 / 0.559992 = 1.71074

\subsubsection{Multiple simultaneous beams: figure of merit}  \label{sec:multbeams}

To simulate the fraction of astrophysical pulses that would be accepted by the ``simultaneous beams'' algorithm, we used the same Monte Carlo study that we performed for DM repetition.  The test pulses were accepted or rejected using the criteria from the ``simultaneous beams'' RFI rejection method.

Monte Carlo statistics, for the same pulses as in Section~\ref{sec:dmrepfom}, are listed in Table~\ref{tab:fom}.

%total pulses = 37572
%
%pulses passing the multi beams test = 37420
%
%fraction passing multi beams = 0.995954
%
%Pulses in workunits analyzed so far for rfi:
%
%total pulses = 256085
%
%pulses passing the multi beams test = 220573
%
%fraction passing = 0.861327
%
%figure of merit = 0.995954 / 0.861327 = 1.1563

\subsubsection{Two simultaneous polarizations: figure of merit} \label{sec:twopols}

If all detected astrophysical pulses were completely unpolarized, or were above threshold in both polarizations, then all of them would pass the ``simultaneous polarizations'' test.  However, even if the astrophysical component of the pulse is unpolarized, the noise component is independent of the astrophysical component.  Thus, pulses near threshold may be detectable in only one polarization.

To simulate the fraction of unpolarized astrophysical pulses that would be accepted by this test, we generated pulse area values with a cumulative distribution $c(s) \propto s^{-3/2}$, or a probability density function $h(s) \propto s^{-5/2}$, where $s$, drawn from the random variable $S$, is the pulse area.  
% To generate this distribution, we take the $-2/3$ power of a uniform distribution.  That is, if $X$ is uniform with distribution $f(x)$, and $S = s(X)$ is the pulse area, then:

% \begin{eqnarray}
% h(s) & \propto & s^{-5/2} \\
% h(s(x)) ds/dx & = & f(x) = \mathrm{ constant} \\
% ds/dx & \propto & s^{5/2} \\
% s^{-5/2} ds & \propto & dx \\
% s^{-3/2} & \propto & x \\
% s & \propto & x^{-2/3}
% \end{eqnarray}

The reason for the $s^{-3/2}$ cumulative distribution is that if we assume a
standard candle source (same luminosity vs. time for all sources), then the
sources at distance $r$ have flux at Earth proportional to $\frac{1}{r^2}$.  The
number of sources within distance $r$ (hence with flux greater than $S \propto
\frac{1}{r^2}$), is proportional to $r^3 \propto S^{-3/2}$.

After determining the test pulse's area, we generate two mini workunit files that
contain the pulse.  Each file combines the pulse with noise randomly, so that different noise is generated in the two mini workunits.
Then, we dedisperse the two files and find the noise-modified pulse areas.
The pulse passes the ``simultaneous polarization'' test if it is above the detection
threshold in both polarizations.  If it is only above threshold in one polarization,
it fails the test.  And if it is below threshold in both polarizations, it would not
be detected at all, so it does not pass or fail.

Note that for a given pulse area threshold, there is a unique probability density function (pdf) with $h(s) \propto s^{-5/2}$, so there is no ambiguity about normalization.  If more astrophysical sources are present, the total number of sources detected will increase, but the pdf will not change.

After $1{,}000$ pairs of test pulses, the figure of merit statistics are given in Table~\ref{tab:fom}. 
The $x$ statistic is not the number of pulses generated ($2{,}000$), but the number detected; some pulses were below threshold.  In the table, the $x$ statistic counts pairs of corresponding pulses as two, whereas the $y$ statistic counts each pair as a single pulse, and $z$ is their ratio.  The $z_2$ statistic was arrived at in a similar manner.

%total pulses inserted = 2000
%
%pulses detected in both pols = 522 $\cdot$ 2
%
%detected in both, after removing 50\% (redundant pulses) = 522
%
%pulses detected in only one pol = 42
%
%total pulses detected (counting redundant) = 522 $\cdot$ 2 + 42 = 1086
%
%fraction of detected test pulses passing multi pols = 522 / 1086 = 0.4807
%
%fraction of pulses in database passing multi pols (after removing 50\%) = 0.03862
%
%figure of merit = 0.4807 / 0.03862 = 12.45

\subsubsection{Frequency profile: figure of merit}

Using the same mini workunits generated for the polarization test, we determine whether the pulse would pass the frequency profile test.  The pulse passes the frequency profile test if its spectrum is flat.  Again, the pdf of the pulse area is unique, given the pulse area threshold, therefore there is no ambiguity as to the pulse powers we should use.

After a Monte Carlo using threshold log\_prob $>$ -1, and after $1{,}000$ pairs of test pulses, the figure of merit statistics are given in Table~\ref{tab:fom}.  
The $x$ statistic is not the number of pulses generated ($2{,}000$), but the number detected; some pulses were below threshold.

%total pulses inserted =        2000
%
%pulses detected =       1075
%
%passing the chi square test =           1075 - 218 = 857
%
%fraction passing the chi square test =                  0.7972
%
%Pulses in workunits analyzed so far for rfi:
%
%total pulses = 246870
%
%passing the chi square test = 149277
%
%fraction passing the chi square test = 0.6047
%
%figure of merit = 0.7972 / 0.6047 = 1.318

\subsubsection{Overall: figure of merit} \label{sec:overallfigmerit}

There seems to be no reason that the astrophysical pulses' passing fractions, as described above,
should be correlated.  (Especially if we exclude the multi-beams test, which is probably unreliable.)
An astrophysical pulse that passes the DM repetition test is no likelier than
any other to pass the multi-pols test, the fraction blanked test, or the
frequency profile test. 

To see this, one has to consider the tests in pairs, and think about the
nature of the tests.  In each case, the property measured by one test is entirely unrelated to
the property measured by the other.  A pulse passes the multi-pols test
if it is strong and/or unpolarized, and it fails the DM repetition test if nearby (noise or RFI)
pulses have the same DM as the signal.  It passes the fraction blanked test if its workunit has a lot of RFI
that overwhelms the receiver or IF electronics, and it passes the frequency profile test if its spectrum is
flat.
So we will assume that the four tests are statistically independent, an assumption which we discuss in the following section.

So we expect the fraction of astrophysical pulses passing all tests
to be: $0.557 \cdot 0.958 \cdot 0.481 \cdot 0.797 = 0.205$, where we have
just multiplied the fraction passing from each test above.  

On the other hand, the fraction of candidate pulses passing all tests
is: $47 / 412001 = 0.000114$, as of February 2010.  (Not counting pulses from our observation of the Crab pulsar.)  This makes for a figure of
merit equal to $1797$, substantially larger than the product of the individual
figures of merit, which is $33$ (see Table~\ref{tab:fom}.)  This makes sense, because the multi-pols test is
designed to catch noise, whereas the other tests are designed to catch RFI.  So we
might expect each test to be less effective on its own,
but more effective in combination with other tests.  (For instance, imagine a fictitious data
set in which $49$\% of all signals are noise, $49$\% are RFI, and $2$\% are real.  If algorithm A
removes all noise, and algorithm B removes all RFI, then the two together have a figure of merit
of $1 / 0.02 = 50$, whereas separately they have $1 / 0.51 \approx 2$.)

\begin{table}
\caption[Figures of merit]{Figures of merit for RFI mitigation algorithms.  In the table, $x$ is the number of pulses analyzed in a Monte Carlo test or other simulation, $y$ is the number of those pulses that pass this test, and $z$ is the fraction passing for simulated pulses.  $x_2$ is the number of candidate pulses analyzed so far, $y_2$ is the number of candidate pulses passing, and $z_2$ is the fraction passing for candidate pulses.  The figure of merit (FoM) is defined as $z / z_2$.  Note that in the row for ``fraction blanked,'' $x$ and $y$ refer to un-blanked space in all workunits (before and after the test is applied), in units of full workunit lengths.}
\vspace{\baselineskip}
\begin{tabular}{lccccccc}
\hline\hline
algorithm & $x$ & $y$ & $z$ & $x_2$ & $y_2$ & $z_2$ & FoM \\
\hline
fraction blanked & $2{,}457{,}187$ & $1{,}368{,}632$ & $0.557$ &
    $256{,}085$ & $122{,}627$ & $0.479$ & $1.16$ \\
DM repetition & $37{,}572$ & $35{,}994$ & $0.958$ & 
    $204{,}994$ & $114{,}795$ & $0.560$ & $1.71$ \\
multi-beams & $37{,}572$ & $37{,}420$ & $0.996$ &
    $256{,}085$ & $220{,}573$ & $0.861$ & $1.16$ \\
multi-pols & $1{,}086$ & $522$ & $0.481$ & 
    & & $0.0386$ & $12.5$ \\
frequency profile & $1{,}075$ & $857$ & $0.797$ & 
    $246{,}870$ & $149{,}277$ & $0.605$ & $1.32$  \\
\hline
\end{tabular}
 \label{tab:fom}
\end{table}

\subsubsection{Statistical independence of our tests} \label{sec:independence}

In Section~\ref{sec:rfimethods}, we have considered several RFI and noise rejection methods.  Our four post-detection methods were as follows:

\begin{enumerate}[A.]
\item Fraction blanked restriction
\item DM repetition
\item Two simultaneous polarizations
\item Frequency profile
\end{enumerate}

We also described how we have tested the figure of merit for each of
these four methods individually, using Monte Carlo simulations.  In
each case, we attempted to make reasonable assumptions about the
distribution of astrophysical and spurious signals that would be
received; these assumptions are detailed in the relevant sections
above.  In the case of methods (C) and (D), we were required to model the power
of the test pulse relative to the noise.  For methods (A) and (B) it was not necessary
to model the power, since these methods depended only on the arrival time and dispersion measure of the pulse.  For each of the four RFI and noise rejection
methods, we have computed, using a simulation, the fraction of
astrophysical pulses that would pass the test; see Table~\ref{tab:fom}.

However, ultimately we are not interested in the individual
algorithms, but in the overall effect of our entire pipeline.  One
might be concerned that astrophysical pulses will pass individual
algorithms, but will fail when they are presented with multiple
algorithms.  Given the four individual passing fractions just
mentioned, can we derive the overall passing fraction for
astrophysical pulses?

To answer this question, we must determine whether our RFI rejection
methods are statistically independent.  That is, if method (A) allows
astrophysical signals to pass with probability $p_A$, and method (B)
allows signals to pass with probability $p_B$, then the two methods
are statistically independent if the probabilities for the joint
outcomes are as given in Table~\ref{tab:statindep}.

\begin{table}[ht]
\caption{Criterion for the statistical independence of two RFI mitigation methods.  Two methods are independent if their outcomes on astrophysical pulses are statistically uncorrelated, so that the probability of each outcome pair is given by the expression in the table.}
\vspace{\baselineskip}
\centering
\begin{tabular}{c c} % centered columns (2 columns)
\hline\hline
Outcome & Probability \\ [0.5ex]
\hline
pass A and pass B & $p_A \cdot p_B$ \\
pass A and fail B & $p_A \cdot (1 - p_B)$ \\
fail A and pass B & $(1 - p_A) \cdot p_B$ \\
fail A and fail B & $(1 - p_A) \cdot (1 - p_B)$ \\ [1ex]
\hline
\end{tabular}
\label{tab:statindep} % is used to refer this table in the text
\end{table}

Another way of saying this is that two methods are independent if
their outcomes are uncorrelated.  In the event that all four methods
are jointly independent, then the joint passing fraction can be found
by multiplying the individual passing fractions.

However, if all methods are not independent, the joint passing
fraction may not be the same as the product of the individual passing
fractions.  In particular, we would like to make sure that the joint
passing fraction for astrophysical pulses is not too low.  Therefore,
we should attempt to verify that the four methods are statistically
independent.  We might attempt to accomplish this by simulating the
four methods using computer generated fake ``astrophysical'' signals.
Unfortunately, it turns out that the statistical independence of the
four methods is highly contingent on our choice of a model for the
behavior of the astrophysical pulses and of the RFI.  We do not
observe the astrophysical pulses directly, nor do we measure all
properties of the RFI.  Therefore, we must make some assumptions about
their properties, and different assumptions will lead to different
models of the four RFI rejection methods.

There is no way out of this: with incomplete information about the
astrophysical pulses and RFI, we must make some assumptions.  We would
not be able to run a simulation to test these assumptions, since any
such simulation would depend upon the very assumptions it was supposed
to test.  Therefore, the best we can do is to carefully state the
assumptions that lead us to model the four methods as being
statistically independent.  To do this, we consider the four methods
in pairs, and explain what it would mean to assume that each pair is 
statistically independent.

When considering the pairs, we are only concerned about factors that
would produce negative correlations, where pulses which pass one test
are likely to fail another.  These negative correlations would reduce
the number of astrophysical pulses that would make it through our
algorithms.  (Positive correlations, where pulses which pass one test
are likely to pass another, would actually improve our ability to
detect astrophysical pulses.)

As we consider pairs of methods, our language will be somewhat repetitive, since each method will be discussed three times.  The reader may wish to examine just one or two entries, to get a flavor of the kind of reasoning that is required.

(A) and (B): fraction blanked and DM repetition:

Astrophysical pulses will fail method (A) if the client blanker detects
a large amount of RFI at a nearby time.  They will fail method (B) if
nearby pulses have the same dispersion measure as the astrophysical
pulse.  This would be most likely to happen if the RFI over a
particular period of time was concentrated at a particular dispersion
measure.  Thus, the two methods will be negatively correlated if
repeating dispersion measures in the RFI are correlated with low
amounts of RFI, and nonrepeating dispersion measures are correlated
with high amounts of RFI.  We are unaware of any such effect.

(A) and (C): fraction blanked and simultaneous polarizations

Astrophysical pulses will fail method (A) if the client blanker detects
a large amount of RFI at a nearby time.  They will fail method (C) if
they are too weak to show up in both polarizations, or if they are
highly polarized.  The two methods will be negatively correlated if
weak or polarized astrophysical pulses are more likely to arrive
during periods of few RFI detections.  Since astrophysical pulses come
from a different source than RFI, this seems unlikely.

(A) and (D): fraction blanked and frequency profile

Astrophysical pulses will fail method (A) if the client blanker detects
a large amount of RFI at a nearby time.  They will fail method (D) if
they are not broadband, or (more likely) if their profile is distorted
by noise or RFI to appear as if it is not broadband.  The two methods
will be negatively correlated if narrowband RFI, of a sort that could
distort an astrophysical signal's frequency profile, is correlated
with low amounts of RFI.  We are unaware of any such effect.

(B) and (C): DM repetition and simultaneous polarizations

Astrophysical pulses will fail method (B) if nearby pulses have the same
dispersion measure as the astrophysical pulse.  They will fail method
(C) if they are too weak to show up in both polarizations, or if they
are highly polarized.  The two methods will be negatively correlated
if weak or polarized astrophysical pulses are less likely to arrive during periods
of RFI with repeating dispersion measure.  Since astrophysical pulses
come from a different source than RFI, this seems unlikely.

(B) and (D): DM repetition and frequency profile

Astrophysical pulses will fail method (B) if nearby pulses have the same
dispersion measure as the astrophysical pulse.  They will fail method
(D) if they are not broadband, or (more likely) if their profile is
distorted by noise or RFI to appear as if it is not broadband.  The
two methods will be negatively correlated if RFI with repeating
dispersion measure is correlated with low amounts of narrowband RFI, of a sort that could distort an astrophysical signal's frequency profile.
We are unaware of any such effect.

(C) and (D): simultaneous polarizations and frequency profile

Astrophysical pulses will fail method (C) if they are too weak to show
up in both polarizations, or if they are highly polarized.  They will
fail method (D) if they are not broadband, or (more likely) if their
profile is distorted by noise or RFI to appear as if it is not
broadband.  The two methods will be negatively correlated if weak or
polarized astrophysical pulses are less likely to arrive at the same
time as narrowband RFI.  Since astrophysical pulses come from a
different source than RFI, this seems unlikely.

Note that these six pairs do not exhaust all of the theoretical
possibilities; it is possible for three quantities to be statistically
dependent even if two are independent; however, similar sorts of
reasoning apply to the case of three quantities' simultaneous
correlations.

\subsubsection{Detection of giant pulses from the Crab pulsar} \label{sec:crab}

The reasoning in Section~\ref{sec:independence} aimed to demonstrate that astrophysical pulses have a good chance to get through our RFI mitigation algorithms.  Additional supporting evidence comes from our detection of several giant pulses from the Crab.

We observed the Crab pulsar intermittently from 11:39 AST to 13:27 AST on June 7, 2009.  We detected $1{,}404$ candidate pulses over that time period, of which $89$ passed all of our RFI and noise mitigation tests.  All $89$ pulses had dispersion measures ranging from $50.0 \pccm$ to $62.6 \pccm$, with a standard deviation of $1.8 \pccm$.  This tells us that Astropulse can typically estimate a dispersion measure to within $1.8 \pccm$, or within $6 \pccm$ in the worst case.  Most of these pulses seem to be duplicate representations of a single astrophysical pulse (or group of pulses).  Astropulse may detect the same pulse in different beams, polarizations, or scales.  The $89$ pulses passing our tests corresponded to $3$ groups clustered in time, indicating that we detected at least $3$ giant pulses.

Of the $1{,}404$ candidate pulses, $171$ had dispersion measures ranging from $50.0 \pccm$ to $62.6 \pccm$.  These candidate pulses clustered into $10$ groups.  We can use this fact to estimate the passing fraction of astrophysical pulses from the Crab.  At most $10$ pulses were astrophysical, therefore the passing fraction was at least $3 / 10 = 30 \%$, which corresponds roughly to our simulated value of $0.205$.

We detected these same pulses independently, using a spectrometer at Arecibo Observatory, and verified that the two methods found the pulses at precisely the same times.  Therefore, we can be confident that we have detected giant pulses from the Crab pulsar, indicating that Astropulse can detect at least some astrophysical signals.

\subsubsection{Limitations} \label{sec:limitations}

As mentioned above, we will not be able to claim a detection based on a single event at or near threshold.  For such a pulse, it would be impossible to know if it was astrophysical or not; we expect to detect four noise pulses that pass the multi-polarization test.  Therefore, any verification would require corroborating evidence.  One type of corroborating evidence would be detection of a distribution of multiple sources.  By examining the properties of these sources (location on the sky, dispersion measure, pulse area, and so forth) we hope to notice patterns that would be uncharacteristic of RFI or noise.  For example, noise would be uniform over the sky whereas an astrophysical source may not be.  Another type of corroborating evidence would be an observation of the pulse at different wavelengths -- for instance, optical transients or gamma ray flares. 

Alternatively, we could improve the multi-polarization test by passing a pair of pulses only if the DMs are similar; this has the potential to substantially cut down the number of false positives due to noise.  If necessary, we could also increase our thresholds further.  Both of these possibilities could potentially be explored in future work.  However, although we may be able to rule out 100\% of noise by these methods, it is difficult to be completely certain that 100\% of RFI was removed.  Therefore, a detection of multiple pulses, which may show some pattern that is uncharacteristic of noise or RFI, would be more valid than detection of a single pulse.

We also note that our passing fraction of $\sim$20\% for astrophysical pulses effectively reduces the survey time by a factor of five, compared to the nominal sensitivity described in Section~\ref{sec:detection}.  We are unable to determine to what extent, if any, this issue affects the other surveys listed in Table~\ref{tab:survey}.

\section{Conclusion}

We designed software to search for microsecond transient radio pulses, using coherent dedispersion to examine the microsecond timescale.  We have focused on evaporating primordial black holes as a potential source of microsecond pulses, but other sources are possible, including giant pulses from pulsars, RRATs, and as yet unknown astrophysical phenomena.
In order to obtain the computational power required for coherent dedispersion, we distributed the software to around $500{,}000$ volunteers, and processed $1{,}540$ hours of observation time at the Arecibo telescope using the Arecibo L-band Feed Array (ALFA) receiver.  We observed simultaneously with each of $7$ beams and $2$ polarizations per beam, for a total of $21{,}600$ hours of data.

In this paper, we have presented the design of the Astropulse experiment, including the scientific rationale and goals of the project, details of the algorithm, data processing, and distributed computing methodology, and our techniques for RFI and noise mitigation.  We compared Astropulse's sensitivity to other searches by computing the minimum detectable pulse area, in $\jyus$, for a microsecond pulse that could be detected by each survey.  In this respect, Astropulse is more sensitive than all other radio surveys we considered, except for that of \citet{den08}.  Astropulse is able to detect pulses of area $54 \jyus$, whereas \citet{den08} can detect pulses of area $8.5 \jyus$.  Astropulse also had the second-largest amount observation time, a total of $10{,}800$ hours of data counting one polarization, whereas \citet{mcl06} observed for $20{,}800$ hours.

We employed multiple techniques to mitigate radio frequency interference (RFI).  Several layers of hardware and software were devoted to blanking known terrestrial radio sources in the vicinity of the Arecibo telescope.  We ruled out signals which appeared at the same DM repeatedly at multiple points on the sky, since an astrophysical source would be located at a single point.  We also ruled out signals which did not occur in both polarizations simultaneously, searching only for unpolarized signals.  Although we lost the opportunity to detect polarized signals, we found that this polarization criterion allowed us to reject a substantial amount of noise.  Finally, we required signals to be broadband, having a similar power at all frequencies, a condition that would hold true for any signal with a short intrinsic timescale.  By combining these criteria, we were able to reject all but approximately $1$ in $10{,}000$ detected pulses.  By performing simulations, we estimated that about $1$ in $5$ astrophysical pulses would pass our RFI rejection criteria.  Therefore, as long as we have detected $5$ or more astrophysical pulses, we would expect some to remain after the RFI mitigation step.  

In a future paper, we will describe the results of the Astropulse experiment, outlining the characteristics of the pulses that remained, and our conclusions about them.

\clearpage

\acknowledgments
    
We are grateful to the National Science Foundation (Grant AST-0808175), NASA (Grant NNX09AN69G), and the Friends of
SETI@home for funding Astropulse and the SETI@home servers. The Arecibo Observatory
is the principal facility of the National Astronomy and Ionosphere Center, which is
operated by the Cornell University under a cooperative agreement with the National Science Foundation.
We would like to thank Luke Kelley for his work on the software radar blanker.
JV would like to thank his wife, Lucy, for her support and encouragement.  
PBD is a Jansky Fellow of the National Radio Astronomy Observatory.

\bibliographystyle{apj}

% \bibliography{$HOME/papers/josh/bib/apj-jour,$HOME/papers/josh/bib/vonkorff}

\clearpage

%\begin{figure}[htp]
%\centering
%\includegraphics[width=140mm]{f3.eps}
%\caption[Blanking attempt]{Blanking attempt.  The image depicts a time vs. frequency ``waterfall plot.''  Frequency is offset so that the $0$ Hz point on the y-axis corrsponds to $1{,}420$ MHz.  The DC signal comes from radar, and the lighter periodicity (along the top and bottom edge of the plot) is an artifact of the hardware blanking.}
%\label{fig:blankingattempt}
%\end{figure}

\end{document}